\documentclass[12pt,preprint]{aastex}

\usepackage{graphicx} 
\usepackage{epsfig}
\usepackage[dvips]{color}
\usepackage{lscape,enumerate}
\usepackage{textcomp}

\newcommand{\chandra}{\mbox{\it Chandra}}

\newcommand{\nh}{\mbox{$N_{\rm H}$}}

\newcommand{\skipthis}[1]{}

\newcommand{\persqcm}{\rm \,cm^{-2}}

\def\micron{\hbox{$\mu$m}}
\newcommand{\be}{\begin{equation}}
\newcommand{\ee}{\end{equation}}
\newcommand{\e}{et al.\ }

\slugcomment{\today}

\shorttitle{Large Stellar X-Ray Flares}
\shortauthors{McCleary  et al.}

\begin{document}


\title{A Survey of High Contrast Stellar Flares Observed by Chandra}


\author{J.E. McCleary\altaffilmark{1}}
\affil{Harvard--Smithsonian Center for Astrophysics,
       60 Garden Street, Cambridge, MA 02138}

\author{S. J. Wolk}
\affil{Harvard--Smithsonian Center for Astrophysics,
       60 Garden Street, Cambridge, MA 02138}



\altaffiltext{1}{Present address: Department of Astronomy, New Mexico State University, P.O. Box 30001,
MSC 4500, Las Cruces, NM 88003-8001; mccleary@nmsu.edu.}


\begin{abstract}
The X-ray light curves of pre-main sequence stars can show variability
in the form of flares altering a baseline characteristic activity
level; the largest X-ray flares are characterized by a rapid rise to
more than 10 times the characteristic count rate, followed by a slower
quasi-exponential decay. Analysis of these high-contrast X-ray flares
enables the study of the innermost magnetic fields of pre-main
sequence stars.  We have scanned the ANCHORS database of Chandra
observations of star-forming regions to extend the study of flare
events on pre-main sequence stars both in sky coverage and in volume.
We developed a sample of 30 high-contrast flares out of the 14,000
stars of various ages and masses available in ANCHORS at the start of 
our study.  Applying methods of time-resolved spectral analysis, we obtain 
the temperatures, confining magnetic field strengths, and loop lengths of 
these bright, energetic flares. The results of the flare analysis are compared 
to the 2MASS and Spitzer data available for the stars in our sample. We find 
that the longest flare loop lengths (of order several stellar radii) are only 
seen on stars whose IR data indicates the presence of disks.  This suggests 
that the longest flares may stretch all the way to the disk. 
Such long flares tend to be more tenuous than the other large flares studied. 
A wide range of loop lengths are observed, indicating that different types of
flares may occur on disked young stellar objects. 

\end{abstract}





\section{Introduction}

Though almost all magnetically active stars show X-ray variability
(e.g. Favata \& Micela 2003; G\"udel 2004), pre-main sequence (PMS)
stars are especially variable.  In ROSAT observations, the X-ray flux
of many young stellar objects (YSOs) were seen to change by factors
of 2 or more between observations separated by less than an hour.
With the onset of nearly continuous multi-day observations by the
XMM-Newton and Chandra X-ray Observatories, rotational modulation of X-ray emitting coronal
plasma has been observed to cause X-ray variability (Marino \e 2003; 
Flaccomio \e 2005).  However, most variability appears to be stochastic in nature: stars' turbulent
magnetic fields induce solar-analog coronal events which heat the
coronal plasma to temperatures of order 10$^6$ -- 10$^8$ K and produce
the bulk of observed X-ray variability (e.g., G\"udel 2004 and references
therein).  This stochastic variability takes the form of flares
altering a baseline characteristic X-ray activity level (Wolk \e 2005)
which itself may be the result continuous of low-level flare activity.

In analogy with the micro-flare heating mechanism proposed for
the solar corona (Hudson 1991), several authors have proposed
that the characteristic X-ray activity level observed in the
light curves of PMS consists of a large number of overlapping,
small-scale flares (e.g. Drake \e 2000; Favata \& Micela
2003; Caramazza \e 2007).  Only intense flares produce
enough flux to be individually resolved against the	
characteristic level; the large majority of flares release
only small amounts of energy, and only the integrated and
time-averaged X-ray emission (Audard \e 2000; G\"udel et
al. 2003) is detected.    

For the purposes of this study, a flare is defined as a sudden
rise in X-ray luminosity accompanied by an increase in plasma
temperature, followed by a slower, quasi-exponential
decay back down towards the characteristic levels of both plasma
temperature and X-ray flux.  The light curve of a flaring star may then 
be divided into characteristic, peak flare, and decay
components corresponding to stages in the evolution of a flare
event on a star.  Analysis of the decay segments of flare
events has become a standard tool to determine the size of the
flaring structure (and so infer other quantities such as the
confining magnetic field).  The analysis is based on
the hydrodynamic modeling of flares by Serio \e (1991),
who derived a thermodynamic decay time applicable to both
solar and stellar flares.  Reale \e (1997) incorporated
this work with information on residual heating from flares'
density-temperature diagrams (Jakimiec \e 1992; Sylwester
\e 1993) to derive a general expression for the flares'
half-loop lengths.  The equations developed by Reale et
al. are essentially inversions of those in Serio \e with a
correction for sustained plasma heating, without which the
loop length may be overestimated.

Application of the model of Reale \e to samples of stars
has yielded a rich dataset.  Taking advantage of the
unprecedented length and depth of the Chandra Orion Ultradeep Project
(COUP), a 13-day observation of the Orion Nebula Cluster,
Favata \e (2005) applied the flare model to a sample of
approximately 30 events on very young stars representing the
most powerful 1\% of COUP flares.  They found flares with
temperatures exceeding 100 MK and loops as long as 0.2 AU,
possibly indicating a direct connection between the star's
photosphere and the inner edge of any existing accretion disk.
Such flares would have a substantial impact on the disk's thermal,
chemical, and dynamical states (see Glassgold \e 2000;
Wolk \e 2005). 

 In contrast, Getman \e (2008a, b) used a modified version of this
modeling technique using the median energy instead of the spectrally fitted 
energies.  Starting with the entire COUP sample they
identified 216 source which experienced flux
increased by at least a factor of four over their
characteristic levels. They found the flares to contain extended,
hot and powerful coronal structures.  Comparison with solar
scaling laws indicated that proposed solar-stellar
power-temperature and duration-temperature relations may not
adequately fit COUP flares.  In the analysis of Getman
\e the hottest COUP flares are found to be brighter but shorter
than cooler flares. They found no evidence that any flare is
produced in star-disk magnetic loops, but instead all are consistent with
enhanced solar long-duration events with both foot-points
anchored in the stellar photosphere. 

Further probing the COUP dataset, Aarnio \e (2010) construct spectral 
energy distributions from 0.3 to 8~$\micron$ for the high-contrast 
flares characterized by Favata \e (2005) and model them to determine 
whether there is enough circumstellar disk material to allow star-disk 
interaction.  Aarnio \e conclude that 58\% of the stars in the COUP 
high-contrast flare sample have no disk material within reach of the 
confining magnetic loops and so conclude that high-contrast X-ray 
flares in general are purely stellar in origin with no footprint in 
the circumstellar accretion disk.

Prior studies were limited in both sky coverage and kinds of
stars examined.  This study will take advantage of the ANCHORS
database of Chandra observations of star-forming regions to
extend the study of flare events both in sky coverage and in
distance.  By not limiting our sample by cluster, age, or
spectral type, we increase the number of flare events studied
and subsequently the strength of any statements about their
properties.

In this study, we will search for stars with flare events in
the ANCHORS database (excluding sources already examined in
the COUP survey), model the flares based on the work of Reale
\e and examine the results of flare modeling in the
context of available infrared data.  The remainder of this
paper will be structured as follows: in \S2 we describe the
selection and basic properties of the samples, in \S 3 we
describe the modeling applied to the flare events and \S 4
contains the results of this modeling.  Discussion of the
results of modeling, including the roles that accretion disks
play in flare behavior, appears in \S 5 and conclusions are presented in \S 6.    

\section{Data}


\subsection {X-Ray Data}

All flares analyzed in this paper are taken from
ANCHORS\footnote{\small{http://cxc.harvard.edu/ANCHORS}}, an archive of
observations of regions of star formation by the \chandra~Advanced CCD
Imaging Spectrometer (ACIS).  The goal of ANCHORS is to provide a
uniform database of X-ray data from different stellar clusters, making
it ideal for this study.  The ANCHORS database is freely accessible to 
the public and includes position, net counts, count rates, hardness ratios, 
and the results of several spectral model fittings for over 14,000 stars 
from over 150 distinct \chandra~observations.\footnote{\small{These values 
are from summer 2008 when this project began.}}  The ANCHORS page for each 
source also contains an evaluation of the credibility of the source based on 
net counts and degree of pileup.  While data from the COUP survey are 
included in the ANCHORS database,with one exception (COUP 1246) this work 
deliberately excludes sources studied in COUP due to their extensive coverage 
in the literature 
(cf. Favata \e 2005, Getman \e 2008a, Getman \e 2008b, Aarnio \e 2010).
 
The X-ray data presented in ANCHORS are standard \chandra~data system pipeline 
level-2 products downloaded from the archive.  Source detection is performed 
using WavDetect in an iterative manner with spatial bins of 1, 2 and 4 pixels.  
WavDetect is run in three bands: 0.5 $-$ 2 keV, 3.0 $-$ 7.5 keV, and 0.5 
$-$ 7.5 keV.  The resulting source lists are then merged.  
Photons with energies below 0.5 keV and above 7.5 keV were filtered out
to minimize contamination from detector background and hard, non-stellar, 
sources.  At each WavDetect source position an extraction ellipse containing 
95\% of encircled energy was calculated based on \chandra~point-spread 
functions.  Background for each source is calculated using an annular 
elliptical region with the same center, orientation and eccentricity as the 
source ellipse.  The inner radius of the background annulus is three times 
the source radius, and the outer radius is six times the source radius.  Net 
source counts are estimated by subtracting background counts in the source 
ellipse from total counts contained in the source ellipse, and multiplying 
the result by 1.053 to correct for the use of a 95\% encircled energy radius.  

After net counts for each source are calculated, spectral, temporal, and 
cross-correlation analyses are performed on the reduced data products.  
The components of the spectroscopy pipeline applied to ANCHORS sources, 
collectively known as YAXX (Yet Another X-Ray Extractor)
\footnote{\small{http://cxc.harvard.edu/contrib/yaxx/}}, are also freely accessible to the public.

Two different timing analyses are applied to all sources in
the ANCHORS database.  First, a Bayesian analysis for
constancy (described in detail by Scargle 1998) splits the
light curves of the stars into periods of constant flux at a
given significance level.  A second timing analysis follows
the method of Gregory and Loredo (GL; 1992), which uses
maximum-likelihood statistics and evaluates a large number of
possible break points against the prediction of constancy. The
Gregory-Loredo method evaluates the probability that the
source was variable and estimates the constant intervals
within the observing window.  The Bayesian method requires a
``prior'' for comparison and hence only detects variability
above a given threshold, usually 95, 99 or 99.9\%.  The
Gregory-Loredo timing analysis returns a probability of
variability as an output. The results of the Gregory-Loredo timing 
analysis are  consistent with the
Bayesian timing analysis.  The Gregory-Loredo analysis
has the added advantage of assigning to each source a
numerical grade from 1 to 10 based on the variability of the
constant-flux blocks.  A variability grade of 0 indicates
constant flux, while a grade of 10 indicates extreme
variability in the count rate.  

To build a sample for flare analysis, the ANCHORS data were filtered
by counts and variability.  The process of selecting sources for flare
analysis began in June 2008, which means that observations must have
been completed by June 2007 to be included in this study.  We find that
sources with high Gregory-Loredo variability grades tend to show more frequently the
archetypal flaring behavior of impulsive rise in X-ray luminosity
followed by gradual decay to characteristic levels.  A variability
grade greater than or equal to 8 ($>$ 99.9\% confidence level) was the
first selection applied to the data.  This first selection yielded
about 2000 sources.  To ensure the extraction of a quality spectrum,
from these sources we chose observations with at least 1000 counts per
source, of which at least 200 come from the peak flare segment of the
light curve. We found this count level to be a good balance between
inclusion of the maximum number of stars for this study and a minimum
number of counts needed to extract a quality spectrum for each segment
of the light curve.  As a further cut, all sources whose light curve did not
show an archetypal flare's fast rise and slow decay were 
discarded for the purposes of this analysis.  A light curve with a
slow rise or no decay could indicate either a flare caused by
different underlying mechanisms and thus not compatible with the
procedure developed in \S 3, or altogether different types of
variability, e.g., rotational modulation.  These last selections
reduced the data set to 29 stars meeting all criteria for counts
and variability.  Finally, all sources were checked for photon pile-up.  
For each source, the highest count rate per pixel after correcting
for \chandra~off-axis PSF is compared against the PIMMS\footnote{\small{http://cxc.harvard.edu/toolkit/pimms.jsp}} 
expected pile-up rate.  None of the sources selected for further study
suffered from photon pile-up greater than 5\%.
Table 1 lists the X-ray properties of the 29 non-COUP stars
involved in this study.  The 17 non-COUP \chandra~observations 
used in this study are summarized in Table 2.  While cluster membership 
is not explicitly determined in this study, the existence of 2MASS 
counterparts for all X-ray sources included and the filtering out of very 
soft or hard X-ray photons limit the likelihood of contamination by non-stellar 
objects for sources in our sample.  The archetypal fast rise and slow decay flaring behavior in 
the sources' X-ray light curves is strong additional evidence for their youth.  Cluster membership 
as displayed in Table 2 is then very highly probable.   

The selections that culminated in Tables 1 and 2 do not
generate an unbiased sample: they specifically select only 
the brightest flares because only the brightest
and longest-lasting flares will have the statistics
necessary for time resolved spectral analysis.  Moreover,
because luminosity scales with the square of
distance, flares from more distant clusters are more
intrinsically intense than those in nearby clusters.  For example, 
flares on sources in M 17 (2100 pc away) must about 100 times more 
luminous than those in the Serpens Cloud Core (260 pc away; though perhaps a bit
further, see Winston \e 2010) to be time-resolved with the
same statistical certainty.  Therefore, the flare
characteristics derived in this paper should be understood as
representative of only the most intense flares present in
stellar coronae, not the ``average'' coronal flares that occur
far more often and make up the characteristic activity level.  
	
\subsection {Infrared Data}
	
Favata \e (2005) argue that for some of the larger flares, the
magnetic loop may have connected the star to the disk.  To 
identify the stars in our sample with disks, the X-ray data 
are supplemented by infrared photometry from the 2MASS All-Sky Point Source
Catalog and Spitzer Space Telescope's IRAC and MIPS
instruments.  The photometry used here is taken from the
Spitzer Young Cluster Survey (SYC; Gutermuth \e 2009), a
systematic IRAC and MIPS 24 micron imaging and photometric
survey of 36 star-forming clusters in which Gutermuth et
al. identified and classified over 2,500 YSOs using
established mid-infrared color-based methods (Winston et
al. 2007; Gutermuth \e 2009).  

Infrared data provide a robust means of detecting disks and
envelopes around young stars via excess thermal emission and
so can be used to identify the X-ray sources in our sample as
protostars (Class I), disked/classical T Tauri stars (cTTs;
Class II), or naked/weak T Tauri stars (wTTs; Class III).  The
T Tauri subclasses may be understood in the framework of a
 disk evolution model whereby Class~I protostars become
Class~II cTTs as the infall process ends. Class~II sources
evolve into Class~III wTTs when the circumstellar disks 
materiel agglomerates into millimeter-sized grains, is accreted 
onto the host star, or is otherwise cleared\footnote{\small{Direct transition from
Class~I to Class~III is not precluded.}}.
	
\begin{figure}
  \centering
  \epsfig{file=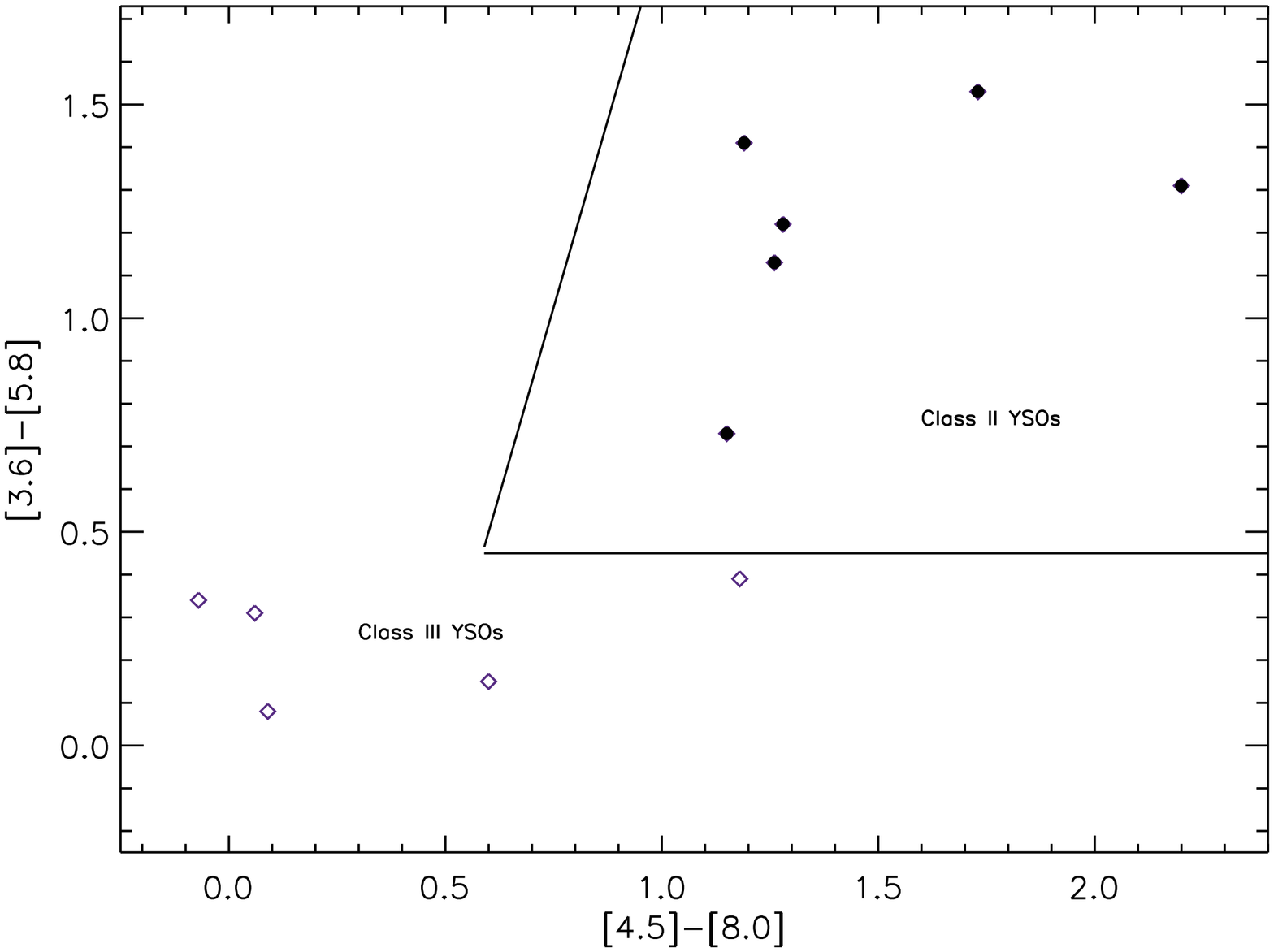,width=0.75\linewidth,angle=90,clip=}
  \caption{{\small [3.6] - [5.8] vs. [4.5] - [8.0] color-color diagram for
  the data listed in Table 3. }}
  \label{fig:IRAC}
\end{figure}

In order to classify the sources in our sample, available IR
photometry from 2MASS, IRAC and MIPS (Table 3) was examined via 
a slew of near- and mid-infrared color-color diagrams.
Gutermuth et al (2005; 2008b; 2009) have shown that the [3.6]
$-$ [5.8] vs. [4.5] $-$ [8.0] color-color diagram robustly
classifies YSOs as either Class II or Class III, as well as
filtering out background contaminants.  The equations
developed by Gutermuth \e (2009) for the detection of YSOs
in the Spitzer Young Cluster Survey were applied to the 15
sources in our sample with IRAC photometry. Figure~\ref{fig:IRAC} shows the
[3.6] $-$ [5.8] vs. [4.5] $-$ [8.0] color-color diagram for
the data listed in Table 3.  Stars with the following color
excesses: 
\be
[3.6] - [4.5] > 0.7
\ee
and
\be	
[4.5] - [5.8] > 0.7
\ee 
are considered Class I YSOs because of their particularly flat or rising spectral
energy distributions.  Two sources in our sample,
CXOANC J162727.0-244049 and CXOANC J034359.7+321403, were
identified as Class I YSOs using these constraints.  

\begin{figure}
  \centering
  \begin{tabular}{cc}
    \epsfig{file=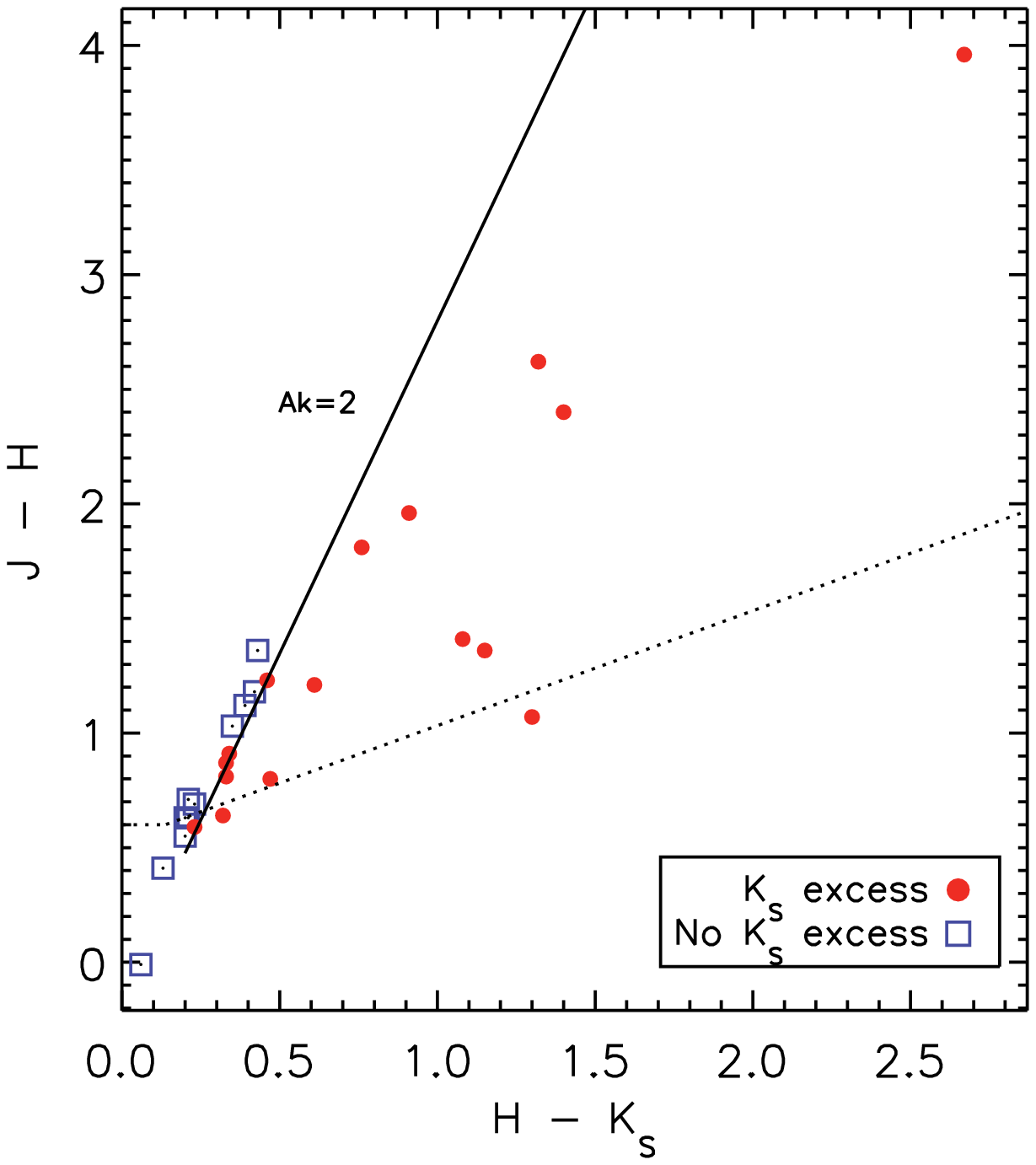,width=0.5\linewidth,clip=} &
    \epsfig{file=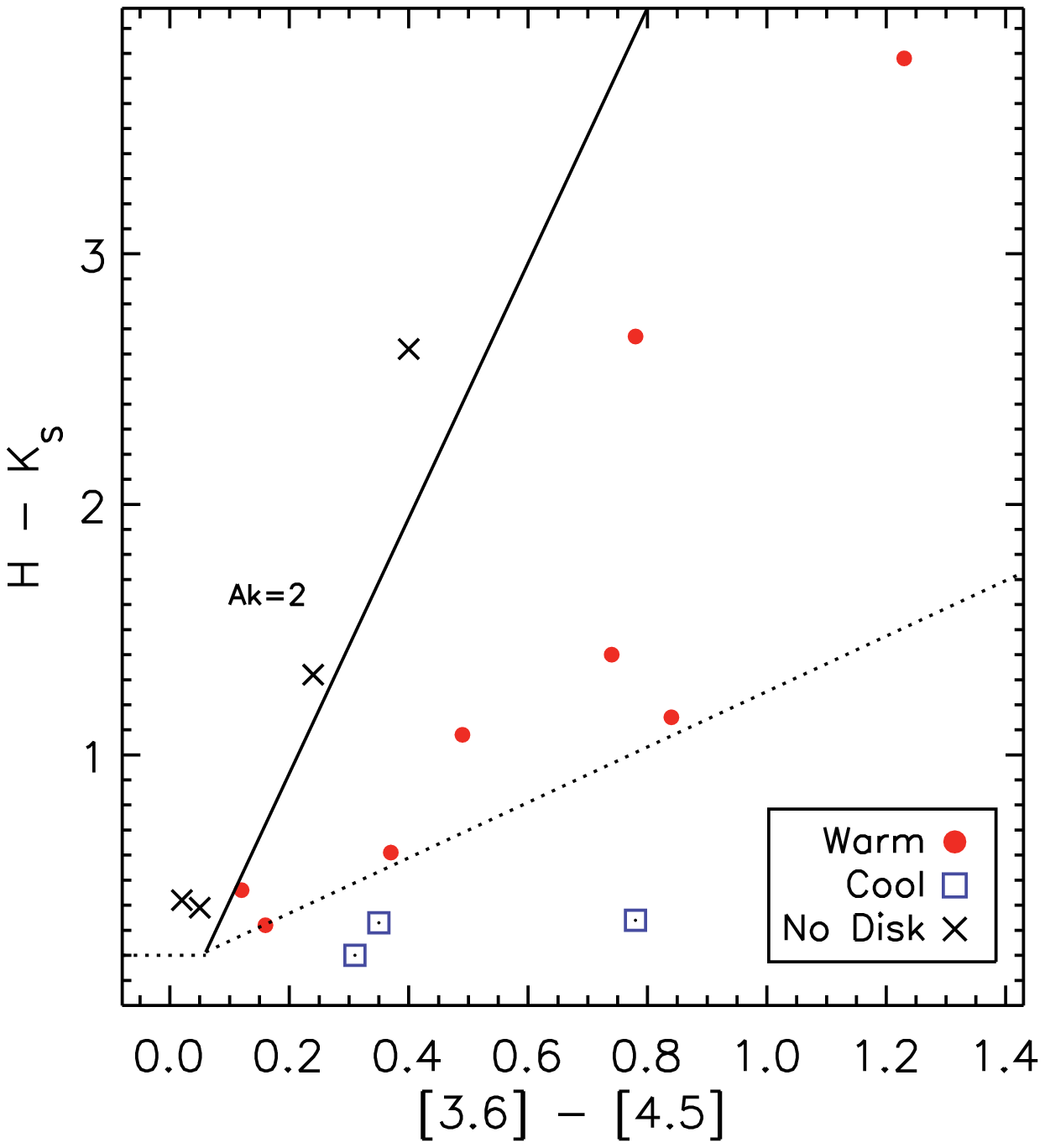,width=0.5\linewidth,clip=} \\
\end{tabular}
\caption{{\small 2MASS band color-color diagrams for the data listed in Table
  3 indicating the presence of a warm circumstellar disks. }}
\label{2MASS}
\end{figure}

The J $-$ H vs. H $-$ $K_s$ diagram shown in Figure~\ref{2MASS} detects warm, 
massive disks that can be easily distinguished from the host star because they are optically thick at K-band.  
The J $-$ H vs. H $-$ $K_s$ diagram does not detect cooler disks that are not optically thick at K-band and thus 
have less contrast against the host star.  The H - $K_s$ vs. [3.6] - [4.5] diagram also shown in Figure~\ref{2MASS}
detects massive but slightly cooler disks that may not be evident in the J $-$ H vs. H $-$ $K_s$ diagram.  Warm disks may be ascribed to stars obeying all of the following constraints (Gutermuth \e 2009): 
\be
for~ [H-K] \le 0.14, [J-H] =0.6
\ee
\be for~  [H-K] > 0.14, [J-H] =0.58 \times [H-K]+0.52\ee
\be for~ [[3.6]-[4.5]]\le 0.06, [H-K] =0.2\ee
\be for~ [[3.6]-[4.5]] > 0.06, [H-K] =1.33\times [[3.6]-[4.5]]+0.133\ee

In addition to six sources already classified with the IRAC 4-band
plot, the J $-$ H vs. H $-$ $K_s$ diagram indicates that four additional YSOs possess warm disks.  
Ages less than 3 Myr for the stars in the host clusters are consistent with the presence of a disk.  
Both the J $-$ H vs. H $-$ $K_s$ and H $-$ $K_s$ vs. [3.6]$-$ [4.5] diagrams are included in Figure~\ref{2MASS}.  Cold gas and dust at a distance of about 3 to 5 AU from the star emits brightly at 24\micron~ (Muzerolle \e 2004).  Accordingly, excess emission ($>$1.5 mag) in the [5.8] $-$ [24] and [4.5] $-$ [24] colors without associated excesses at shorter wavelengths is generally attributed to the absence of an inner disk and identifies stars with disks that are cold (i.e. transition disks), although other configurations 
such as rings are possible as well (Calvet \e 2008).  Source CXOANC
J032929.2+311834 showed excess 24~\micron~ emission and no
corresponding excess in either the [4.5] $-$ [5.8] or [3.6] $-$ [4.5]
colors and so is identified as having a transition disk. 
Further YSO classifications were obtained from literature searches for
six of the ten sources whose IR photometry did not conclusively place
them in an evolutionary class.  Several studies were consulted for these data 
(Flaccomio \e 2006 for NGC~2264 -- obsid 2540, Giardino \e\ 2008 for NGC 752 -- obsid 3752; 
Kohno \e 2002 for Mon R2 -- obsid 1882; Damiani \e 2004 for NGC 6530 -- obsid 977; Dahm \& Hillenbrand 2007 for NGC 2362 and M8  -- obsids 4469 and
3754 respectively).  
	
\section{Flare Modeling}
\subsection{The Hydrodynamic Model of Stellar Flares}
Modeling the spectra for each source is a multi-step procedure guided
by the principle that flares are isolated events periodically altering
an underlying characteristic spectrum -- even if the underlying
spectrum is composed of multiple unresolved flares.  Because the
coronae of stars in our dataset are unresolved, the key to
determining flare characteristics is good modelling of the spectra
of the quiescent, peak and decay segments of the flare.  The model
itself has evolved over about 20 years, beginning with the
thermodynamic timescale for flare loop decay of Serio \e (1991) as
a function of loop half-length $L$ (considered from one of the flare
footpoints to its apex), peak temperature $T_{PK}$, and
e-fold decay time $\tau_{TH}$:

\be
\tau_{TH}=3.7\times10^{-4} {L\over{\sqrt{T_{PK}}} }  
\ee
In principle, if $T_{PK}$ and $\tau_{TH}$ can be determined, equation 7
can be inverted to obtain the flare loop length.  However, the
quantities $\tau_{TH}$ and $T_{PK}$ 
cannot be directly observed.  Instead, they must be derived based on
other observable quantities from the data.  

Because the archetypal flares' rise time is significantly shorter than the
decay time, $\tau_{TH}$ in equation 7 assumes impulsive heating concentrated
only at the beginning of the flare event.  If the heating is
not strictly impulsive, the flare plasma may be subject to
prolonged heating extending into the decay phase.  As a
result, the flare's e-fold decay time measured directly from the light curve
($\tau_{LC}$) would be longer than the flare's intrinsic decay
time ($\tau_{TH}$) and application of equation 7 would
result in the overestimation of flare loop length.  The slope 
of the flare plotted in a log T vs. log $n_e$ space ($\zeta$) can
be used as a quantitative measure of the timescale of any
sustained heating.  Reale \e (1997) show that $\zeta$ provides a
diagnostic of the ratio between intrinsic ($\tau_{TH}$) and observed
decay times ($\tau_{LC}$) such that:  
\be
{\tau_{LC} \over {\tau_{TH}}} = F(\zeta)
\ee
The actual form of $F(\zeta)$ is observation-specific, depending on
the band-pass and spectral response of the instrument used to obtain
flare data.  For values of $\zeta$ between 0.32 and 1.5, the ratio between observed
and intrinsic flare times for ACIS observations has been calibrated to:
	\be
	{\tau_{LC} \over {\tau_{TH}}} = F(\zeta)={0.63 \over {\zeta-0.32}}+1.41
	\ee
Equation 9 allows for the substitution of $\tau_{TH}$ in Equation
7 with the observable quantities $\tau_{LC}$ and $\zeta$.  Favata \e also
calibrated the actual peak flare temperature, $T_{PK}$, as a function
of the observed temperature (in kelvins), $T_{OBS}$, for ACIS observations:   
	\be
	 T_{PK}=0.068\times T_{OBS}^{1.2}  
	\ee
 With corrections for sustained heating and instrumental
 response in place, the equation for flare loop half-length is
 based entirely on observable quantities and becomes: 
	\be
	L={\tau_{LC}\sqrt{T_{PK}}\over 3.7\times 10^{-4} F(\zeta)}
	\ee
		
The loop length, in turn, may be used to determine the volume of
the flare loop.  If the ratio $\beta$ between the flare's
radius and its length is fixed at the solar value of $\beta
\sim 0.1$, then the volume of the flare loop may be derived
using equation (8) from Favata \e: 
	\be
		V=2\pi\beta^2L^3
	\ee
Consequently, we can infer the flare's resulting plasma
density in cm$^{-3}$ and confining magnetic field($B$) in
gauss using equations (9) and (10) from Favata \e by
estimating the electron density ($n_e$): 	
	\be
	n_e=\sqrt{{EM\over V}}=\sqrt{{EM\over 2\pi\beta^2L^3}}
	\ee
	\be
	B=\sqrt{(8\pi)(kn_eT_{PK})}
	\ee
where, through the Ideal Gas Law, the quantity $kn_eT$ is
equivalent to the flare loop's plasma pressure.  It should be 
noted that this estimate of the magnetic field is actually a lower
limit, only the minimum field strength required to confine plasma of density
$n_e$ and temperature $T_{PK}$.  The flare's confining magnetic field may be 
higher than inferred from equation 14.  Equations 11,
13, and 14 were applied to all sources in our sample; the
resulting values of $L$, $B$ and $n_e$ are listed in Table 4.
Uncertainty in the value of $\beta$ used to derive $B$ and
$n_e$ introduces considerable systematic error into their reported
values, and so the calculated values for the magnetic field
and the electron density should be understood to represent relative magnitudes.
Assuming an extreme case of a circular flare loop for which
$\beta = 1$, the estimates differ by a factor of 10 for the
electron density and a factor of 3 for the magnetic field.

It should be noted that this formulation for stellar flares is degenerate 
in that it doesn't differentiate between a single long flaring loop and 
an arcade of several smaller flaring loops for a given length of plasma.  
Such complex flaring structures have been observed on the Sun, and they can have
very simple light curves.  Because the sources in our sample are 
unresolved, it is impossible to distinguish between a single flare loop and an arcade
of smaller loops.  

\subsection{Application of Model to X-ray Data}
\begin{figure}
\epsscale{.750}
\plotone{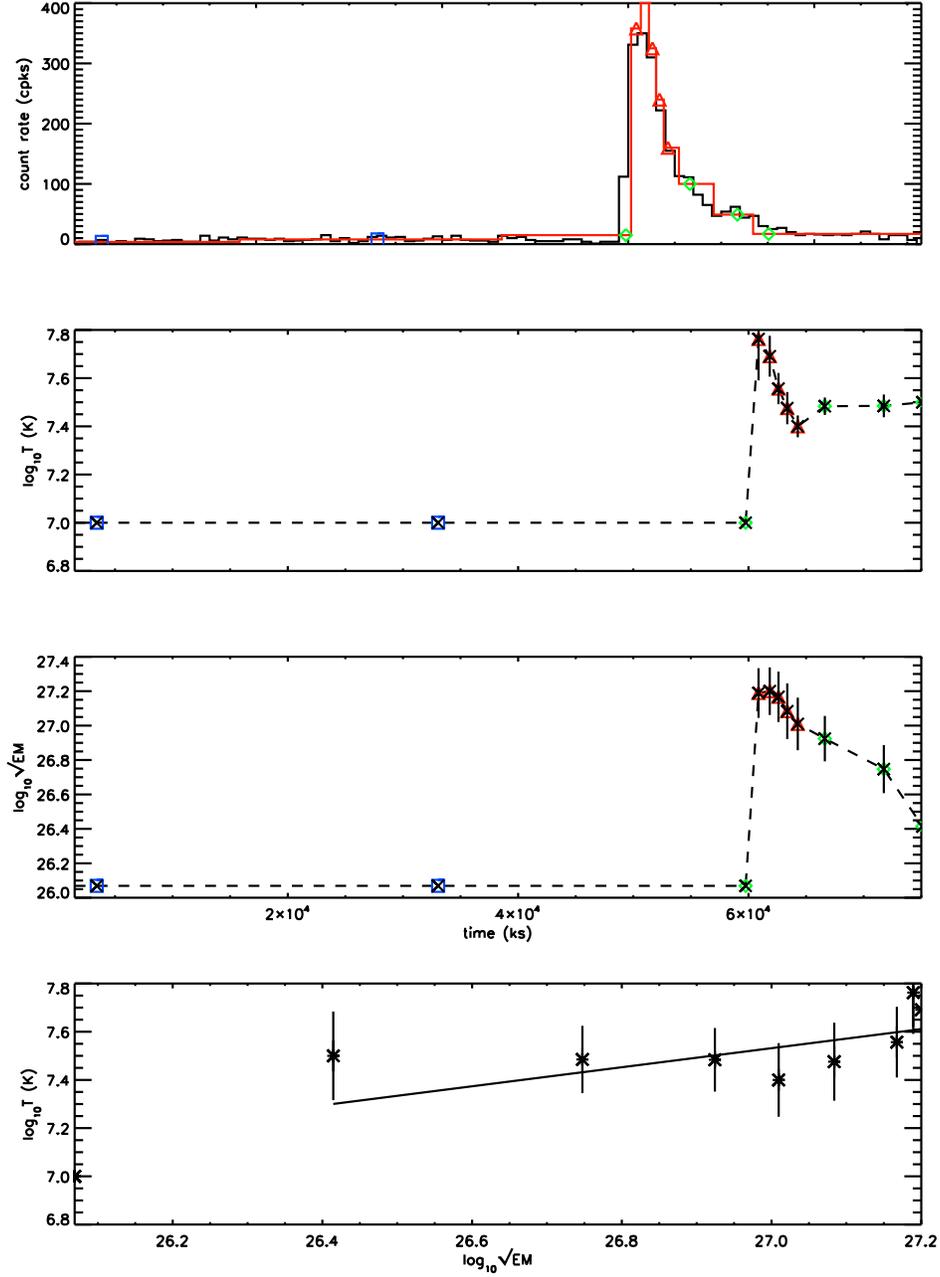}
\caption{{\small  Flare analysis described in
  \S3 as applied to an archetypal flare, CXOANC J182943.0+010207
  ($\chandra$ observation 4479).  The topmost plot is the light curve for the
  source.  Segments considered at the characteristic level are plotted
  as blue squares, rise and decay segments are plotted as green
  diamonds, and peak flare segments are plotted as red triangles.
  The next two plots show the evolution of the flare's temperature and
  emission measure with time.  The bottom plot shows the evolution of the flare in the
  log $T$ - log$\sqrt{EM}$ plane, with the points used to obtain $\zeta$
  connected by a best-fit line (the characteristic point on the lower left is excluded).}}
\label{paper_plots}
\end{figure}

To ascertain $\zeta$ and so characterize the flare loop, resolved spectra of each phase of the flare are required.
In order to obtain these phase-resolved spectra, the light curve of
each source is divided into a characteristic portion, a peak
potion, and a decay portion.  A finer resolution on the flare
translates to a better-constrained value for $\zeta$ and more confident
derivation of flare characteristics, so the peak and decay portions of
the light curve are themselves subdivided into multiple segments.  Light
 curve subdivision relies on timing information from the Bayesian blocks (BBs) generated with
the Scargle (1998) algorithm and a prior consistent with 95\%
confidence that the photon arrival rate had changed.  Depending on the particular 
resolution of the source, there can be as few as three or as many as nine individual segments associated with a
flare event on a star in our sample.  Each segment of a flare event was required to have a minimum of 100 to 150
counts, and in all cases \nh~was fixed to the value reported in the ANCHORS
database, usually of order $10^{22}$ cm$^{-3}$.  This limits uncertainties in temperature to less than
30\%.  We assume that the \nh~is dominated by interstellar gas or material associated with the local disk, 
not material associated with possible coronal mass ejections or accretion streams.  The subdivision of the 
lightcurve into segments is shown for a sample source (\chandra~observation 4479) in the topmost plot of 
Figure~\ref{paper_plots}.

After subdivision, spectra are generated for all flare event segments
with the specextract routine from CXC's CIAO 3.4 software
package.\footnote{\small {http://cxc.harvard.edu/ciao3.4/index.html}} The
resulting spectra are fitted with an absorbed
multi-temperature model using Sherpa 2.0 to obtain the evolution of
temperature and density needed for the calculation of flare
parameters.\footnote{\small {Creating spectra in CIAO 3.4, as we did, and then
  processing those spectra in Sherpa 2.0 (associated with CIAO 4.0)
  does not result in errors.}}   Because Sherpa fits are based on
chi-squared minimization, the spectra are binned to ensure robust
statistics. 

Spectral fits occur in several steps.  First, the characteristic
spectrum is fitted with a three-temperature model with all abundances
freed in order to provide the best possible fit to the data.  The
model assumes the emission spectrum of collisionally ionized diffuse
thermal plasma as calculated with the APEC code (Smith \e 2001).  To
account for the interstellar medium and the circumstellar environment 
of the sources, a multiplicative absorption component (WABS;
Morrison \& McCommon 1982) is included in the model.  Initial guesses
for hydrogen column density came from the ANCHORS estimate.  We assume 
that the gas column is dominated by interstellar gas or material associated 
with the local disk, not material associated with a possible coronal mass 
ejection or possible accretion streams.  The
results of this stage of modeling have limited physical significance,
but generally provide an excellent fit to the data and serve as a
mathematical representation of the characteristic level from which
evolution in density and temperature may be studied.  After a model is
determined for the characteristic spectrum, all parameters associated
with it are frozen.

To fit the peak and decay segments of the flare event with the
frozen characteristic model, we include another component consisting
of a simple one-temperature APEC model with only temperature and
normalization as free parameters.  The frozen portion of this new, 
compound model represents the underlying characteristic level, while
the one-temperature component monitors changes in temperature
and emission measure associated with the flare.  The compound model was 
fitted to each segment of all flare events in our sample.  
Normalizations from the fits are interpreted as emission measures for 
the particular segment of the light curve.  This stage of modeling leads 
to a series of points tracking the evolution of a flare in temperature 
and density (via the proxy of $\sqrt{EM}$).  
The two middle plots of Figure~\ref{paper_plots} show this evolution 
for \chandra~observation 4479.

We then calculate $\zeta$ and subsequently the sustained heating
correction $F(\zeta)$ for all sources in our sample by fitting the slope of $\log
T$ vs. $\log \sqrt{EM}$ (see the bottom of Figure~\ref{paper_plots} for an example).  
Finally, loop lengths and other derived quantities are determined for the flare
events in our sample through application of equations 9 through 14.

It should be noted that our procedure closely follows the
analysis of Favata \e (2005) but does not replicate it exactly:
differences in the sample selection techniques, light curve
subdivision, etc., could potentially affect derived characteristics
for the flare.  In order to calibrate and verify our implementation,
we processed source 1246 from the COUP survey (Favata \e 2005).  We
derived a value of 1.03$\pm$ 0.21 for $\zeta$, which agrees within
uncertainty to the value of 0.90 $\pm$ 0.18 reported by Favata \e
Our loop half-length of $50 \times 10^{10}$ cm (with a range of $43 -
55 \times 10^{10}$ cm) is also consistent  the value of $40 \times
10^{10}$ cm reported by Favata \e ($37 - 47 \times 10^{10}$ cm).   
All other derived flare characteristics agree to within a factor of about 2.
Infrared photometry for COUP 1246 is included in Table 3, and its derived flare 
characteristics are included in Table 4.  

\section{RESULTS} \label{results}

Table 4 summarizes the physical characteristics of the flares inferred
from the analysis in \S 3, and includes the energy
released by the entire flare event and the YSO class of the sources when it
is known.  The total energy released by a flare event was calculated
as follows: for each segment of the flare modeled,
the APEC/WABS model outputs an X-ray flux as well as temperature and
emission measure.  Multiplying this flux by the duration of
the segment and scaling by the distance to the source yields the energy 
emitted by that particular flare segment.  Summing over all segments modeled 
yields the total energy emitted by a flare.  The $log_{10}$ of the average 
energy emitted by all flares in this sample is $35.7 \pm 0.7$ ergs.  The average 
energy emitted by flares on known Class II sources  is $36.4 \pm 0.8$ ergs, while
flares on known Class III YSOs emit an average of  $36.0 \pm 0.5$
ergs.  These two values agree within uncertainties and
we conclude that the presence of a circumstellar disk does not appear
to affect total energy emitted by a flare.   

\begin{figure}
\epsscale{.750}
\plotone{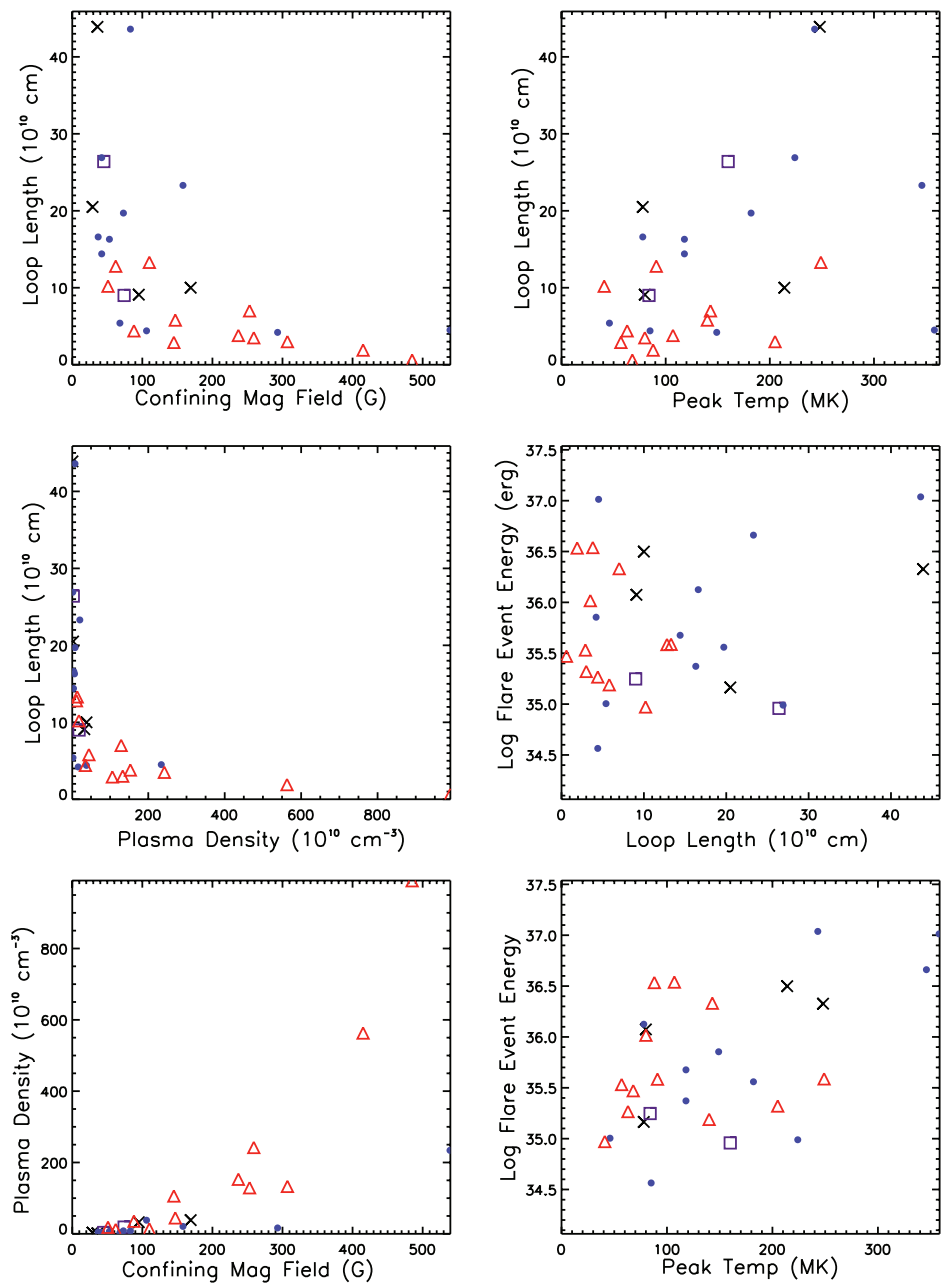}
\caption{{\small Plots of the relationships between the various flare
  characteristics presented in Table 4.  Purple squares represent Class I YSOs, 
blue circles denote Class II YSOs, red triangles represent Class III YSOs,
and unclassified YSOs are shown as black crosses.}}
\label{BasicResults}
\end{figure}

Figure~\ref{BasicResults} shows plots of the relationships among
derived flare characteristics.  Although the longer flares in this
sample tend to be hotter and release more energy, flare temperature
and emitted energy themselves are not strong predictors of loop length
(fig.~\ref{BasicResults}b, ~\ref{BasicResults}d), a result also seen by 
Getman \e (2008a).  As predicted by equation 11, flare loop length is more 
or less proportional to the square root of temperature, but significant scatter 
is introduced by the $F(\zeta)$ parameter measured from the data itself.  Flare 
temperature and emitted energy are weakly correlated, with a Pearson $r$-coefficient 
of $0.52$ (fig.~\ref{BasicResults}f).  Equation 13 predicts that 
plasma density varies with the square of the magnetic field strength, a result 
we confirmed in Figure~\ref{BasicResults}e.  

As seen in Figures~\ref{BasicResults}c and~\ref{BasicResults}e, flare
loop length is anti-correlated with magnetic field strength and plasma
density; the longest flares in this sample are tenuous and weakly
confined.  In fact, Figures~\ref{BasicResults}c and~\ref{BasicResults}e suggest 
that flare loops longer than $1 \times 10^{11}$ cm {\em require} magnetic fields 
weaker than 200 G and plasma density less than $5 \times 10^{11}$ cm$^{-3}$.  
Although this observed dependence of loop length on magnetic field strength and 
plasma density is predicted by equations 11, 13, and 14, the trends shown in 
Figures~\ref{BasicResults}c and~\ref{BasicResults}e are not guaranteed by 
the model: the reported flare properties depend on emission measure and peak 
flare temperature, quantities which are determined from the data and do not depend on the
hydrodynamic model described in this section.  
 
\begin{figure}
\epsscale{.750}
\plotone{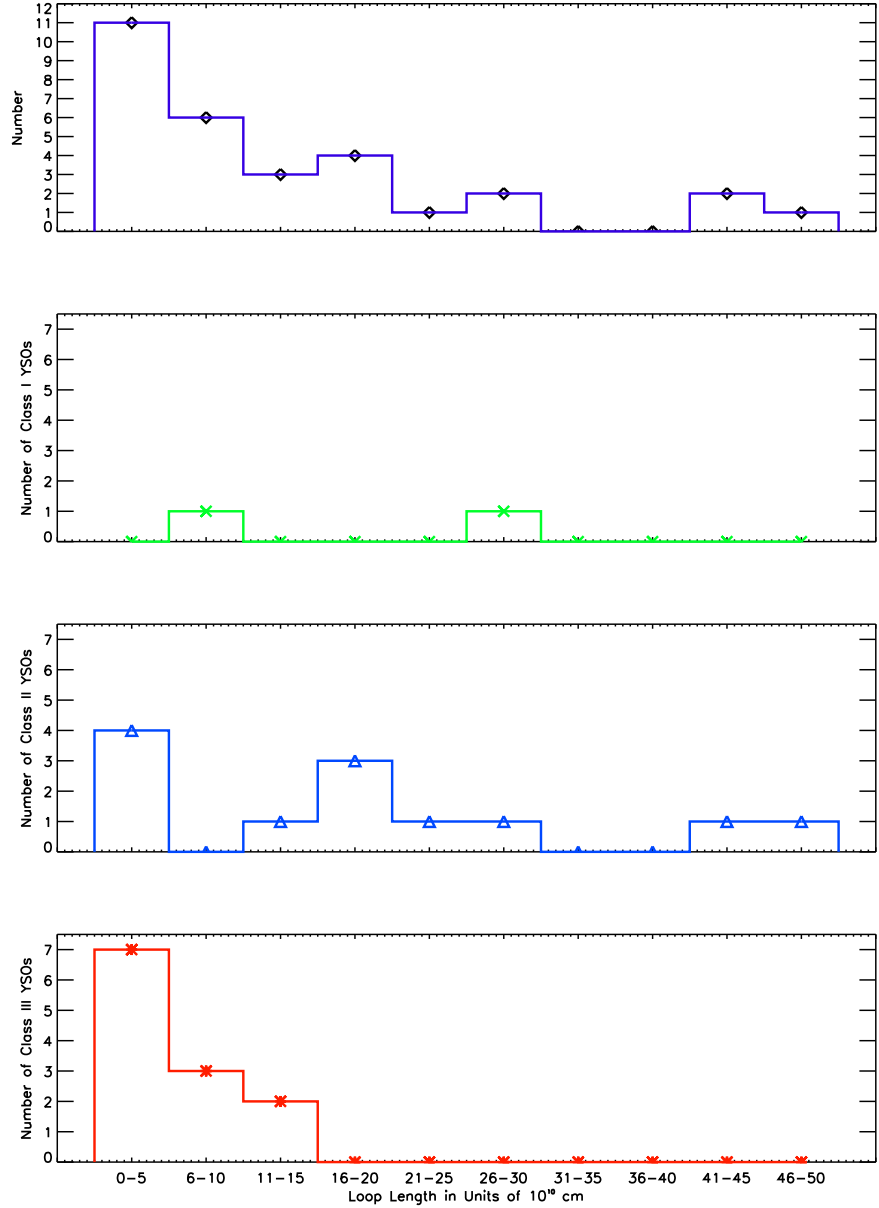}
\caption{{\small Loop length of flares sorted by IR class.  The uppermost plot
shows the distribution of loop lengths for all IR classes, including
unclassified YSOs (represented as black circles).  The Class I YSOs are
represented by green crosses, the Class II YSOs by blue triangles, and
the Class III YSOs by red asterisks.}}
\label{LoopL_hist}
\end{figure}

\begin{figure}
\epsscale{.750}\
\plotone{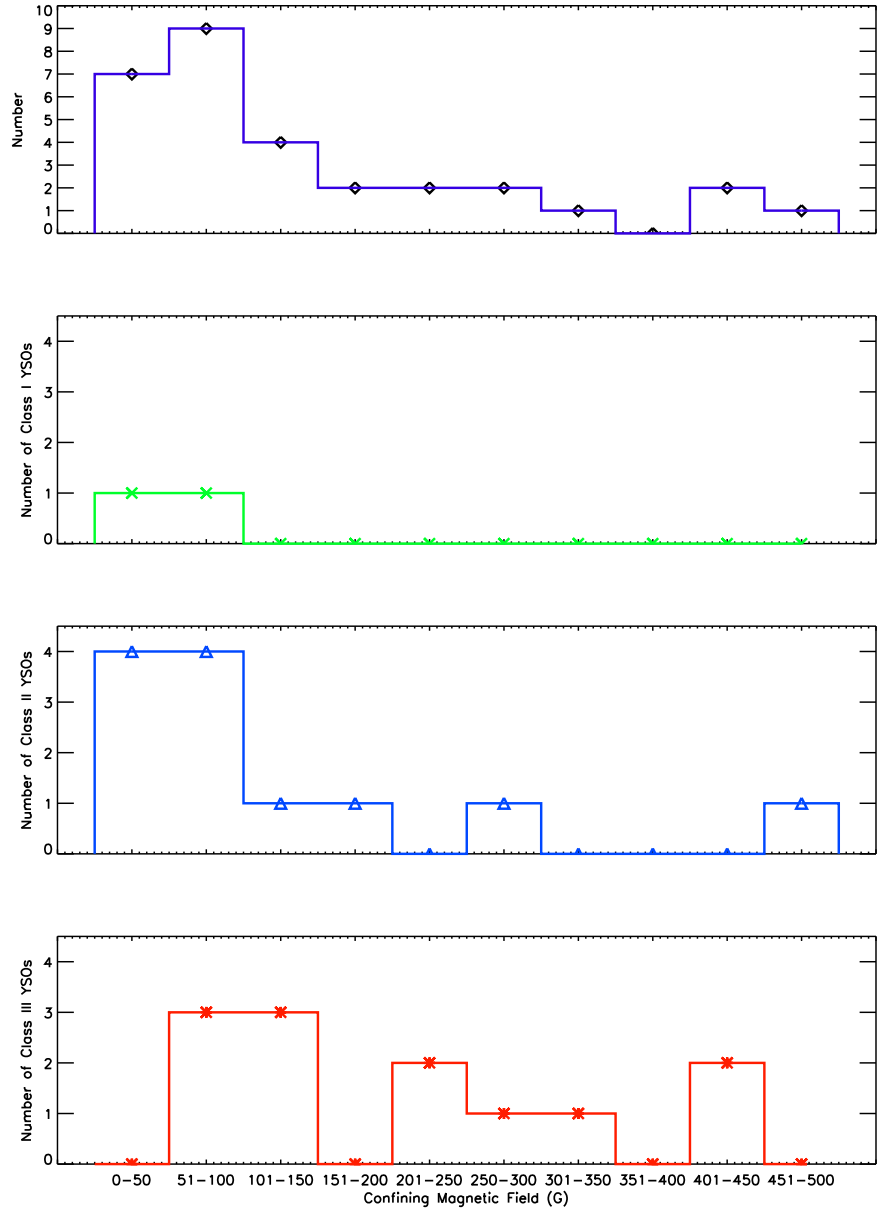}
\caption{{\small Flare confining magnetic field strength sorted by IR class.  In the uppermost
  plot, the distribution of magnetic field strengths for all IR
  classes is shown, including unclassified YSOs.  Colors and symbols are the same as Figure~\ref{LoopL_hist}.}}
\label{B_hist}
\end{figure}

\begin{figure}
\epsscale{.750}\
\plotone{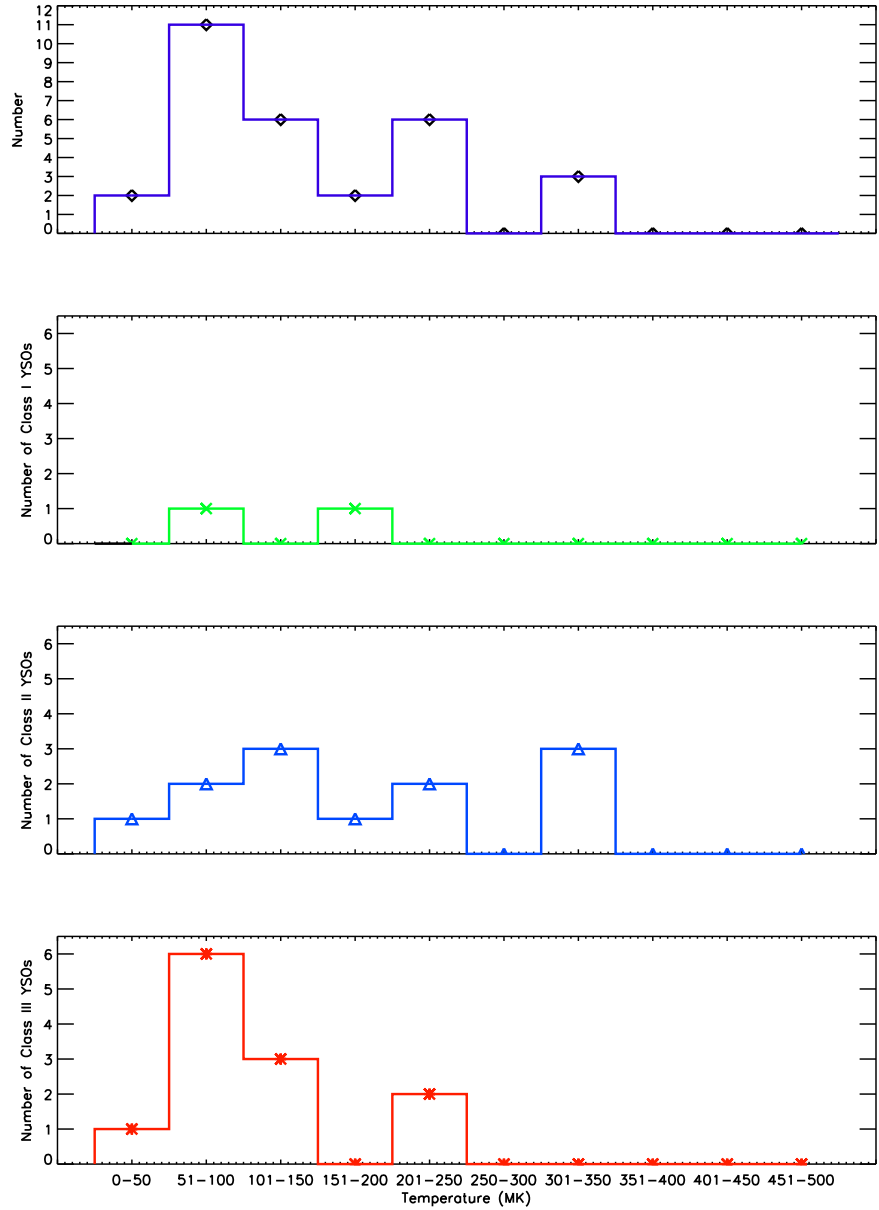}
\caption{{\small Peak flare temperature sorted by IR class.  The uppermost plot
shows the distribution of peak flare temperatures for all IR classes, including
unclassified YSOs.  Colors and symbols are the same as Figure~\ref{LoopL_hist}.}}
\label{T_hist}
\end{figure}

\begin{figure}
\epsscale{.75}\
\plotone{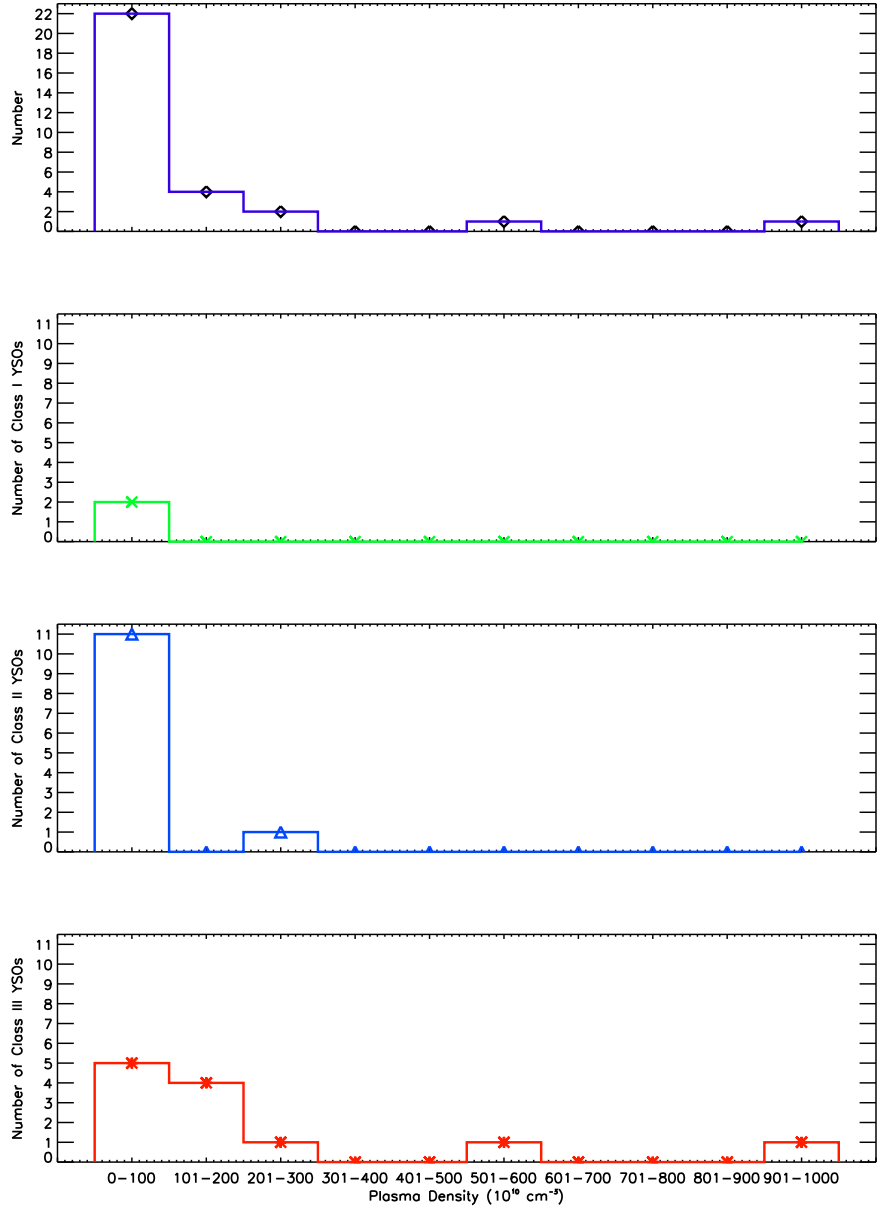}
\caption{{\small Flare plasma density sorted by IR class.  In the uppermost plot, the
  distribution of plasma densities for all IR classes is shown,
  including unclassified YSOs.  Colors and symbols are the same as Figure~\ref{LoopL_hist}.}}
\label{n_e_hist}
\end{figure}

Histograms of derived flare
characteristics segregated by source YSO class
are shown in Figures~\ref{LoopL_hist} through~\ref{n_e_hist}.  
The loop length histogram shows that about two--thirds of the sources
in our sample have loop lengths equal to or smaller than the stellar
radius (assuming that $R_{\star}= 3 R_{\odot}$, cf. Baraffe \e 1998).  About half of the
flares in the sample are ``hot,'' with peak plasma temperatures exceeding
100 MK.  Figures~\ref{B_hist} and ~\ref{n_e_hist} show that about two--thirds of
flare events in our sample are confined in magnetic fields less than
150 G, and about 90\% of the flares have plasma density less than $2
\times 10^{12}$ cm$^{-3}$.

We now examine differences in the flare characteristic histograms by YSO class 
in order to determine the effect of circumstellar disks on flaring behavior.  
Figure~\ref{LoopL_hist} shows that while the distribution of the
twelve identified Class II YSOs' loop lengths is fairly spread out
over all bins out to $5 \times 10^{11}$ cm, none of the eleven
identified Class III YSOs has a loop length longer than $1.3 \times
10^{11}$ cm.  To quantify this result, the distributions of
loop lengths were compared using a two-sided Kolmogorov-Smirnov (KS) tests.
We assumed that the molecular envelopes of material in-falling onto outer parts of a disk
 associated with Class I YSOs would
have no effect on flaring behavior, so the Class~I distribution was combined with the Class II distribution to test
whether the presence of thick circumstellar disks affects flare loop
length.  Comparing the Class~III loop length distribution to the
ensemble of Class I and Class~II loop lengths with a KS test yields
an 11\% probability that the two distributions are drawn from the same
parent population. Though not definitive, this result is suggestive of
a difference between Class I/II and Class III YSOs in the mechanism(s)
driving at least some flare production.    

Kolmogorov-Smirnov tests are also used to quantify the magnetic field
distributions seen in Figure~\ref{B_hist}.   As a general rule, Class I YSOs
 observed in X-rays tend to be hotter than Class II and III YSOs\footnote{
\small{An observation bias exists in that the only Class I flares we can detect 
are those hot enough to penetrate the high nH columns.}}, 
and since magnetic fields associated with flares 
vary as $\sqrt{T}$ (cf. equation 14), we expect the magnetic field
strength distributions to reflect this difference.  The
KS statistic for the magnetic field strength
distributions of identified Class II and Class III YSOs yields a 97\%
probability that the magnetic field strength distributions originate in the same parent
population.  We then compared the combined the Class II and Class III
distributions with the magnetic field distribution of identified Class I
YSOs.  The resulting probability that the Class II/III distribution is
drawn from the same parent population as the Class I distribution is 
only 0.7\%.  Although the Class I distribution comprises only two soures, 
the KS statistic at least suggests
that the magnetic fields of Class I YSOs are quantitatively different
from those on Class II and III YSOs.     

Figures~\ref{T_hist} and~\ref{n_e_hist} show the distribution of
peak flare event temperatures and plasma densities by YSO class.
``Hot'' flares with $T_{PK} > 100$ MK are not observed preferentially
to arise on a particular YSO class; a KS
test on the set of temperatures results in a $98\%$ probability that
the Class II and Class III temperature distributions originate in the
same parent distribution.  However, the two flares in our sample with
$n_e > 30 \times 10^{10}$~$cm^{-3}$ occur on Class III YSOs.  These findings
are consistent with Figure~\ref{BasicResults}b and ~\ref{BasicResults}c, which
indicate that loop length is not a strong predictor of peak
temperature but is rather anti-correlated with plasma density.
Even if lengthened flares are preferentially found on Class II YSOs, the
hottest flares will not necessarily be found on Class II YSOs as
well.
  
As discussed in $\S 2.1$, our sample has a bias towards the
brightest flares on young stellar objects.  It is of interest to
estimate the frequency with which these intense flares occur.  By
multiplying the number of sources in each Chandra observation included
in our sample by the reported exposure time and summing over all
observations, this sample may be estimated to contain approximately
2.9 million ks (i.e. $2.9 \times 10^{9}$ seconds) of ``on star'' time.  
Twenty-nine large flares were observed in this ``on star'' time, which
translates to one large flare per 100,000 ks of observation time.
Phrased differently, the average YSO in our sample produces one of
these intense flares, characterized by fast rise to a ten-fold
increase in flux over the characteristic level followed by a smooth 
quasi-exponential decay, about once every three years.

\section{Discussion}\label{discussion}
\subsection{The Role of Disks in Flare Behavior}
Lengths of the flaring structures in this sample range from $0.53 \times 10^{10}$ cm to $50 \times 10^{10}$~cm,
($\sim 0.03 - 3.0 R_{\odot}$; assuming a fiducial $R_{\star} = 3 R_{\odot}$)\footnote{\small{Current stellar evolution models predict stellar 
radii ranging from 1.5 to 4$R_{\odot}$ for masses
between 0.2 and 3$M_{\odot}$ (Siess \e 2007)}}.  Their confining magnetic fields range between
tens and hundreds of Gauss. Compact flares with $L \le 0.5 R_{\star}$ could easily 
be contained within the YSOs' dipole fields and
appear similar to coronal events seen on ZAMS or MS stars.  Compact
flares have been frequently observed on YSOs, which led to the
historical modeling of YSO magnetic fields as enhanced solar-type
magnetic fields with footpoints anchored in the stellar 
photosphere (cf. Feigelson \& Montmerle 1999 and references therein).
While explaining the compact flares on YSOs, the enhanced solar-type
model cannot account for the long magnetic structures (of order
several stellar radii) implied by the most extended flare events in our sample.
Without an enormous field tension as a counterbalance, the centrifugal
force on magnetic structures several stellar radii long and anchored
solely on the stellar surface would likely be sufficient to rupture the loops
and eject any plasma before flaring. Moreover, for the long loops in our sample, the data do not 
support the existence of the high magnetic field tension expected in a tightly wound arcade of flares.  
In fact, as shown in Figure 4a,  loop length is anti-correlated with confining magnetic field.

  Confronted with the failure of the enhanced
solar-type model to explain long flares, the most probable magnetic
field geometry capable of supporting coherent magnetic fields over
several stellar radii is the connection of the stellar photosphere to
the inner edge of the circumstellar disk via magnetic flux tubes.  

Several theoretical models (e.g., Shu \e 1993, 2000; Feigelson et
al. 2006) predict that PMS stars are magnetically coupled to the inner
edge of their (Keplerian) disks at corotation radii defined as 
\be
R_c=\sqrt[3]{{GM\over\Omega^2_o}}
\ee
Theoretical studies (Uzdensky \e 2002; Matt \& Pudritz 2005) indicate that the presence of a disk truncates 
the stellar magnetosphere so closed magnetic loops cannot extend much further into 
the accretion disk than this corotating inner edge.  The steady state solution dictates that the 
inner edge of the accretion disk rotates at the same rate as the star, but dissipative effects lead 
to differential rotation between the stellar photosphere and the inner edge of the
disk.  This differential rotation induces shearing of the long magnetic field lines connecting the 
star and disk, producing the stressed magnetic field configuration necessary to drive flares (Shu
\e 2000).  

In typical low-mass PMS stars with rotation periods ranging from 2-10 days (cf. Edwards \e 1993), 
equation 15 predicts an $R_C$ of about $3-5 R_\star$ from the stellar surface.  The radius from 
the star at which dust sublimates may be modelled as 
\be
R_s=\frac{\sqrt{Q_R}}{2} \left(\frac{T_\star}{T_s}\right)^2R_\star
\ee
where $T_s$ is the dust sublimation temperature and $Q_R$ is the ratio of the absorption 
efficiencies for dust at the color temperature of the incident and remitted radiation fields
(Tuthill, Monnier \& Danchi 2001).  For a dust sublimation temperature of 1500 K, and a dust 
absorption efficiency $Q_R$ of 1, the sublimation radii range from 2$R_\star$ for a stellar
temperature of 3000 K to 8$R_\star$ for a stellar temperature of 6000 K.  D'Alessio \e (2004) add a term
to equation 16 which accounts for luminosity from any accretion streams.  This extra luminosity
may push the dust sublimation radius up to three times further out from the star, but its inclusion
requires detailed infrared spectra to constrain the radiation field and dust opacities.  As such, we 
do not attempt to estimate the contribution to the stellar radiation field due to accretion streams.
It may be of interest to calculate dust destruction radii for the two flares in our sample in which 
star-disk interaction is the most likely.  For J182016.5-161003, with flare loop length 
$4.4 \times 10^{11}$ cm, there is no reliable radius or temperature estimates available.  
Assuming a Siess model-based $R_{\star} = 3 R_{\odot}$ for a stellar mass between 0.2 and 3$M_{\odot}$,
and a corresponding effective temperature of 4000 K, equation 16 yields an inner disk 
edge of $7.4 \times 10^{11}$ cm from the star, which likely out of reach of the flare.  If a model 
temperature of 3075 K is used, equation 16 yields a dust sublimation radius of $4.5 \times 10^{11}$ cm and flare-disk 
interaction becomes probable.  The case for flare-disk interaction in J182016.5-161003 remains ambiguous.  
For the second long flare, COUP 1246 with flare loop length $5.0 \times 10^{11}$, Favata \e (2005) publish a mass of 0.2 $M_{\odot}$ 
and a stellar radius of 1.6 $R_{\odot}$.  Using a corresponding Siess model temperature of 4000 K, 
the corresponding dust destruction radius is $4.0 \times 10^{11}$ cm, which is easily within reach of the flare.

Five of the 14 flares in our sample associated with disked YSOs have $L \le 0.5 R_\star$, much like the 
flares described by Pandey \& Singh (2008) on active evolved stars.  
Compact flares can be wholly contained by the YSO's dipole field regardless 
of the presence of circumstellar accretion disks; for
this reason compact flares are generally considered analogous to
coronal events seen on ZAMS or MS stars.  
The wide range of loop lengths in the sample suggests that either of two types of flares may
occur on disked YSOs: compact flares -- analogous to flares on evolved
stars -- or long and the result of star-disk magnetic connections.   

Star-disk magnetic coupling can explain the YSO class-specific
distribution of loop lengths seen in Figure~\ref{LoopL_hist}.  
The disks of Class I and II YSOs have inner radii close to 
surface of the central star and are thus far more likely to support the extended magnetic structures
of star-disk coupling than the meager or non-existent disks associated with
Class III YSOs.  If star-disk magnetic coupling drives long flares we should not see any long flares
associated with Class III YSOs.  The loop length histograms of
Figure~\ref{LoopL_hist} are evidence for the existence of star-disk
magnetic coupling, as flares longer than $1.5 \times 10^{11}$ cm are
only observed on Class I and II YSOs. 
  
From these results and the results published in the COUP survey (\S5.2), a consistent picture of star-disk 
magnetic interaction in YSOs emeges.  As part of normal chromospheric activity on a young star, 
the local magnetic field is arranged in arcade loops analagous to those seen on the Sun.  Convection on 
the surface of the star causes shuffling of the loop footprint, which eventually stresses the magnetic 
field loops and causes a reconnection event and flare.  Since YSOs have large surfaces compared to 
Sun-like MS stars, convective activity (and shuffling of the arcade loop footprint) is comparatively reduced.  
If the surface underneath the loop footprint is calm enough, and if they are 
equatorially located, the magnetic loops can grow large enough to interact with the inner edge of disk 
by drawing out ionized material.  The presence or absence of a circumstellar disk plays no role in determining 
high-contrast flare energetics, but should a field line reach out and find disk material, the disk supports 
this extended magnetic structure and the accompanying extended high-contrast flare once the flux tube 
ruptures.  For a detailed discussion of scenarios in which magnetic field loops can grow to several stellar radii, 
see MacNeice \e (2004).  

Such extended high-contrast flares are rare.  There are 4 candidates in our sample of 12 Class II YSOs, 
which suggests that these kinds of flares occur once a decade on a given Class~II YSO.  
However, as we note in \S5.2, our sample is not sensitive to flares with $e$--fold cooling times longer than
about half a day.  Equation 11 implies that the $e$--fold cooling time is proportional to flare loop length,
so our sample is effectively biased towards more compact flares, and the rate of occurence of extended
flares may be higher by a factor of a few.

Twelve of the the thirty flares in our sample occur on disked Class II YSOs, and another twelve are detected  
on Class III YSOs.  The total flare rate on Class II and Class III YSOs is consistent with being the same, 
with no significant observational bias between the two classes.  This is further evidence that the flux flares in 
our sample are part of the normal variety of stellar flares: extended flares simply erupt in such a 
configuration that they can interact with the disk.  Flares that reach the disk are not a separate variety of flare, 
but rather a special case of a more common (if large) flare.

\subsection {Comparison to Other Flare Samples}

The derived flare characteristics presented in Table~4 can be compared to
results for the six late-type active evolved dwarfs studied by Pandey
and Singh (2008), which used a similar technique for flare analysis.  
The six stars in their sample were observed for 30
to 60 ks, during which time Pandey and Singh report a total of 17
flares occurred.  Flares in their sample do not last longer than 10 ks
and usually only last a matter of minutes.  Although the electron densities are similar in
both our samples, the total energy
released by flares in the sample of Pandey and Singh tend to be about
three orders of magnitude smaller than the energy emitted by
flares in our sample. Pandey and Singh conclude that flares in their sample
are analogous to solar arcade flares; the numerous differences between
the flares seen on the active dwarfs and the YSO flares indicate 
that unique influences must be driving the special case of long flares
seen on Class II YSOs. 

A comparison of the derived flare characteristics presented in Table~4
to the COUP survey's flare characteristics (see Favata \e 2005,
Table 1) shows some global differences: the flare e-fold decay times
in this study (mean of 11.0 ks) are significantly shorter than those
reported in the COUP survey (mean of 67 ks); the dependence of derived
flare characteristics on $\tau_{LC}$ leads to smaller loop lengths in our sample
than those reported by Favata \e 
The mean loop length in our sample is $14 \pm 10 \times 10^{10}$
cm and the mean magnetic field is $152 \pm 106$ G compared to $103 \pm 80
\times 10^{10}$ cm and $351 \pm 300 $G respectively in the Favata \e sample. This
discrepancy may be explained in terms of differences in exposure times
between the two studies.  The average effective exposure time in our
sample is 96 ks, with a maximum exposure time of 151 ks, significantly
less than the unequaled 850 ks effective exposure
time in the COUP survey.  Our sample is effectively biased against
flares lasting longer than ~75-100 ks, which restricts $\tau_{LC}$ and
thus the loop length (equation 11).  Consequently, the flares in our
sample tend to be shorter than flares in the COUP sample, which has no such
restriction on $\tau_{LC}$.

In \S4 we estimate the rate of occurrence of high-contrast flares in our sample
to be 1 per 100,000 ks of exposure time.  The COUP sample comprises
1500 stars nominally observed for 850 ks, but the effective COUP
exposure time is actually somewhat longer than 850 ks because flares
may occur during the 14-hour gaps between observations (when $\chandra$ passes through 
the Van Allen belt) and still be detected by their
aftereffects.  With a revised effective exposure time of about 1130
ks, using our metric for rate of occurence we expect 17 high-contrast flare detections in the COUP sample.
However, Favata \e report 32 flare detections.  This difference
can be explained in several ways.  First, the median age of stars in the ONC is
only about 1 Myr, while the median age for stars in our sample is a
slightly more mature 2.5 Myr, with some stars as old as 10 Myr.  The
younger stars in the ONC might simply be more active and produce more high-contrast flares.
A difference in flare selection criteria between our two samples could
also affect the number of flares reported.  Favata \e required only that
three contiguous MLBs (see \S2.1) be significantly elevated above the
characteristic level and that the decay in the log T - log $\sqrt{EM}$
plane be sufficiently regular (i.e. monotonically decreasing).  
Our selection criteria were somewhat more exacting: in addition to requiring a linear decay
in the log T - log $\sqrt{EM}$ plane, we required the archetypal flare
behavior of fast rise and exponential decay with a peak count rate
around 10 times the characteristic level.     

We can also compare our sample of flares to the study published by
Getman \e (2008a,b).  Getman \e constructed a sample of 216 flares
observed during the COUP mission and analyzed them using a different flare
spectral analysis technique (MASME) that avoids nonlinear modeling (Getman
\e 2008a and references therein).  The same hydrodynamic modeling
used in \S3 was used to derive flare properties like loop length
and peak flare temperature.  The selection criteria for
flares in the Getman \e sample were somewhat more relaxed than those
used in this study: all flare morphologies, not just archetypal
fast-rise-slow-decay flares, were accepted as long as the peak count
rate was $\geq 4$ times that of the characteristic level.
Sample selection aside, many of the derived flare
properties are comparable to the results presented in Table~4: Getman
\e report a median $T_{OBS}$ of 63 MK, which is similar to our
median $T_{OBS}$ of 50 MK, and a median loop length of $43 \times
10^{10} \persqcm$ for their sample.  While the median loop
length in our sample is only $10 \times 10^{10} \persqcm$, only two of 
the thirty flares in our sample have lengths equal to or greater than $43
\times 10^{10} \persqcm$; the difference in median loop lengths is
likely due to our smaller sample size.  Finally, Getman \e
conclude that disk presence has no effect on peak flare luminosity
and total flare energy (2008b), a result which we confirm in \S4.  

The principle difference between our study and Getman \e regards the
role of disks in flaring behavior.  As discussed in \S5.1, we find
that long flares of order $R_{C}$ occur only on disked PMS stars and
conclude that in the case of these very long flares, one footpoint of
the magnetic loop must be anchored on the accretion disk.  Getman \e
come to the opposite conclusion, observing flares of order
$R_{C}$ or longer almost exclusively on YSOs categorized as Class III
in their study (2008b).  While Getman \e find little actual variation
in flare properties between disked and diskless stars, they argue that
due to their faster rotation periods, Class III YSOs actually have
significantly smaller co--rotation radii than Class II YSOs and so
regardless of length in cm, flares on Class II YSOs rarely if ever
reach the disk and are anchored on the stellar photosphere alone.   

Finally, we can compare our sample to the study of COUP flares recently published by Aarnio \e (2010).  
Their goal was to understand the role that circumstellar disks play in the energetics of high-contrast 
X-ray flares.  The authors construct spectral energy distributions in the wavelength range ~0.3 $-$ 8~$\micron$ 
(extending in some cases out to 24~$\micron$) and model them to determine whether there is circumstellar disk 
material in sufficient proximity to the flares' confining magnetic loops to allow star-disk interaction.  
If $R_{\star}$, $M_{\star}$ and $T_{eff}$ are supplied, SED modelling can yield disk parameters including mass and 
sublimation radius.  
This technique characterizes the circumstellar disk more quantitatively than the near-infrared color excesses used in this 
study and has the advantage of detecting even cool disks with large inner holes.  Aarnio \e conclude that 58\% of 
the stars in their COUP high-contrast flare sample have no disk material within reach of the confining magnetic 
loops and so argue that high-contrast X-ray flares in general are purely stellar in origin, neither triggered nor 
stabilized by star-disk interactions.  
The authors further argue that as long as the confining magnetic field at the end of the loop is strong 
enough, the loop is stable even without being anchored on the disk.  We agree that disk presence has no effect on flare 
energetics, however our Figure~\ref{BasicResults} shows that confining magnetic field falls off sharply with loop 
length irrespective of disk presence and does not support this scenario.

Aarnio \e do agree that for 10 of objects studied, star-disk interaction is possible and that the 
IR photometry of a further 8 sources in the sample does not permit conclusive statements about disks.  
Of the 15 flares in their sample with loops longer than about 10 $R_{\star}$, 5 flares appear within reach of the inner 
edge of the disk, and another 5 fall into the unknown category and have flare loop lengths of order the 
dust destruction radius.  While not all flares studied by Aarnio \e require star-disk interaction, the 
longest flares in the COUP high-contrast flare sample could be stablized by a circumstellar disk in the scenario 
presented in \S5.1.

\subsection{Comparison to Superflares}
We have already discussed the bias in this sample towards high luminosity
flares; these intense flares on YSOs may be analogs to so-called
``superflares'' seen on magnetically active main sequence stars.
Schaefer \e (2000) apply the term ``superflare'' to flares with
energies ranging from $10^{33}$ to $10^{38}$ ergs on main sequence
stars of spectral class F8 to G8 with no close binary companion and no
evidence for rapid rotation.  

Osten \e (2007) reported the detection of
an X-ray superflare on the active binary system II Pegasi.  The II Peg superflare 
radiated $6 \times 10^{36}$ ergs, the vast majority of which was radiated at energies less than 10 keV 
(see figure 4 of Osten \e 2007).  This value is comparable to the average energy radiated by
flares in our sample, $1.66 \times  10^{36}$ ergs.  It is
thus tempting to draw an analogy between the YSO flares in our sample
and superflares like those reported on II Peg, and so we can compare superflare 
characteristics to our flare sample.  The reported II Peg
superflare temperature is somewhat higher than observed (not peak)
temperatures in our sample: various model fits to II Peg yield flare
temperatures between 118 and 152 MK, consistent with the highest
observed temperature in our sample of 126 $\pm$ 30 MK.  The entire 
II~Peg superflare only lasted for 11 ks, which is closer to the average
e-fold decay time of flare events in our sample, not their entire
duration.  Based on 6.4 keV emission, Osten \e report a loop length range of 0.5 $-$ 1.3
$R_{\star}$ for the superflare, which falls within the range
of loop lengths observed in our sample, 0.2 $-$ 3
$R_{\star}$.  Osten \e use $n_e$ of about $10^{11}$ cm$^{-3}$, consistent with our sample's
median plasma density of $1.2 \times 10^{11}$ cm$^{-3}$.  Using the values for plasma 
density and observed temperature reported by Osten et al., equation 14 can be used to estimate that the 
II Peg superflare is confined by a magnetic field of about 50 G, which is also consistent with the magnetic 
field strengths observed in our sample.  Finally, Osten \e estimate that superflares
like the one on II Pegasi occur once every 5.4 months, considerably
more frequent than the value of one flare every three years calculated
in \S4 for our sample.  Although similar flare characteristics do not guarantee 
identical flare production mechanisms, if II Peg may be considered representative of
superflares on other active MS stars, such superflares seem analogous to the flares in our sample.  

\subsection{Iron 6.4 keV Fluorescence}
Another important observation noted during the II Peg superflare was
the detection of iron K-shell fluorescence emission at 6.4 keV (Osten \e 2007).  
Fluorescence emission has been well documented for solar flares and has also been observed on several
YSOs sources (Tsujimoto \e 2005, Skinner \e 2007, 2009).  The line can penetrate a large
column density (up to $10^{24}$ cm$^{-2}$) and so serves as an
excellent probe of circumstellar environments.  The existence of
Fe K$\alpha$ emission on the Sun is usually attributed to photoexcitation
of iron in the photosphere by bremsstrahlung radiation.  A similar
mechanism is believed to produce the emission on YSOs with two
differences: the fluoresced iron is located in the circumstellar disk,
not the photosphere, and the fluoresced material is irradiated by the
X-rays from magnetic reconnection events, not quasi-stable stellar X-rays.  
Tsujimoto \e report that detection of 6.4 keV emission line is
correlated with larger-than-average flare amplitude, and they predict
that 6.4 keV emission due to photoionization occurs exclusively in
young stars with disks.  The emission was reported in only 6 out of
127 sources examined by Tsujimoto \e  Careful examination of the 30
sources in our sample for the presence of 6.4 keV emission yielded no
positive detection of the line.  Our sample contains only 14 known Class I and II
YSOs with disks thick enough for photoionized 6.4 keV, assuming the
same rate of detection as seen by Tsujimoto \e we would have expected only
about a 50\% probability of detecting an iron fluorescence line from
one of our disked stars.  Thus, zero detections in the 14 disked systems studied is consistent with their detection rate.

\section{Conclusions}\label{conclusions}

In this study, we built a sample of bright flare events on YSOs
suitable for time-resolved spectral analysis from the ANCHORS database.  
We modeled the flare events with the time-dependent hydrodynamic
model of Reale \e (1997).  The results of flare modeling were then
examined in the context of YSO classification information from
available near- and mid-infrared data.  We identify large flares as
having a fast rise in count rate to a level about ten times the
characteristic level, followed by a slower quasi-exponential decay
back down to characteristic level. We calculate that YSOs produce
flares like the ones in our sample about once every three years.  The
flares in this sample exhibit a variety of plasma temperatures and
loop lengths, but the more extended flares are tenuous and confined by weak magnetic
fields. We find that flare loop length is anti-correlated with
magnetic field strength and plasma density and that flare temperature
and energy do not correlate with flare loop length, a result also noted by
 Getman \e (2008b) and Aarnio \e (2010).  Flare properties derived in this study
are comparable in magnitude to those presented by Getman \e (2008a,b) and Aarnio \e (2010)
but significantly smaller than those presented by Favata \e (2005).
This discrepancy is most probably caused by the much longer
observation time of sources in the COUP mission than in our sample,
which allows for the discovery of more extended flares with longer
decay times. 

Comparison of the flares in our sample with a well-studied superflare
on II~Pegasi, a magnetically active main sequence star, reveals that
although the peak flare temperature
Class I and II YSOs, and the lack of loop lengths longer than $13
\times 10^{10}$ cm on the identified Class III YSOs, suggests and energy emitted by the II Peg
superflare is comparable to energies in our sample, the entire II Peg
superflare event was relatively short. 
Overall, the the II~Peg superflare is consistent with the compact
flares in our sample, suggesting analagous flare production mechanisms.  
A search for 6.4 keV fluorescent iron emission as observed in the II Peg superflare 
in our sample yielded a null result.  
This was however, consistent with the number of stars in our sample and the 
generally observed low rate of detection.  

The detection of flares of order several stellar radii in length on
the existence of long magnetic structures connecting the star and
disk.  We suggest the following scenario for star-disk magnetic interaction: equatorially located
magnetic loops like those seen on the Sun can grow large enough to interact with the inner 
edge of a circumstellar accretion disk, provided convection on the chromosphere does not shuffle the loop 
footprint too much.  The presence of a circumstellar disk does not determine flare energetics, but 
should a field line reach out and find disk material, the disk supports this extended magnetic structure.  
When the magnetic loop ruptures due to, e.g., differential rotation between the star and the disk, 
an extended flare results.  Star-disk magnetic coupling thus explains
the distribution of loop lengths seen in Figure~4: the thin or
non-existent disks associated with Class III YSOs cannot support
the coherent magnetic field structures required for long flares.   
A limitation of the hydrodynamic model employed in this study is the inability 
to distinguish between a single large flare loop or an arcade of smaller loops for a 
given loop length L.  However, we find it evidentiary that the only flares with L of order the
dust destruction radius occur on stars with disks, and we do not find
extended flare loop lengths for Class III sources.  While these results are not conclusive proof of star-disk magnetic 
interaction, they are highly suggestive.



\acknowledgments

The authors gratefully acknowledge many useful comments from the anonymous referee, as well as the assistance of Ettore Flaccomio with X-ray data reduction, and the assistance of Robert Gutermuth in sharing IRAC and MIPS data for the sources in our sample.  
This publication makes use of data products from the Two Micron All Sky Survey, which is a joint project of the University of Massachusetts and the Infrared Processing and Analysis Center, funded by the National Aeronautics and Space Administration and the National Science Foundation. J.E.M. was supported by the CXC guest investigator program supported this work through grant AR7-8003A. S.J.W. was supported by NASA contract NAS8-03060.



{\it Facilities:} \facility{2MASS}, \facility{SST (IRAC/MIPS)}, \facility{CXO (ACIS)}.

\clearpage

\clearpage

\clearpage

\clearpage

\clearpage

\clearpage

\clearpage

\clearpage

\begin{deluxetable}{lcrcccrccc}
\tabletypesize{\scriptsize}
\tablecaption{X-ray properties of the 29 non-COUP sources used in this study.\label{tbl-1}}
\tablewidth{0pt}
\tablehead{
\colhead{ANCHORS ID} &\colhead{Off-axis distance}  & 
\colhead{Counts} & \colhead{Counts} & \colhead{Aperture Flux\tablenotemark{a}}  & 
\colhead{Luminosity\tablenotemark{b}} & \colhead{Hardness} & \colhead{Hardness} & \colhead{Hardness}\\
\colhead{~} & & \colhead{~}  &
\colhead{~} & \colhead{~}   & \colhead{~} &
\colhead{Ratio 1} & \colhead{Ratio 2} & \colhead{Ratio 3}\\
\colhead{[CXOANC ...]} & \colhead{[arcsec]} & \colhead{[raw]} & \colhead{[net]} & 
\colhead{~} & \colhead{~} & \colhead{~} & \colhead{~} & \colhead{~}
}
\startdata
J162726.95-244049.8 & 406.0 & 2997 & 2979.0 & -11.02 & 30.11 & 0.97 & 0.48 & 0.99 \\
J162622.49-242252.0 & 1026.0 & 6824 & 6660.3 & -10.60 & 30.59 & 0.92 & 0.75 & 0.99 \\
J162715.46-242639.1 & 464.5 & 1271 & 1258.5 & -11.39 & 29.70 &  0.94 & 0.70 & 0.99 \\
J162624.10-242447.2 & 179.3 & 2246 & 2241.6 & -11.07 & 30.18 &  0.96 & 0.80 & 1.00 \\
J053456.83-051133.0 & 498.1 & 2341 & 2311.8 & -11.52 & 30.59 & -0.46 & 0.27 & 0.21 \\
J053525.73-050949.3 & 253.7 & 1450 & 1444.7 & -11.70 & 30.38 & -0.25 & 0.50 & 0.37 \\
J053629.62-045359.3 & 1248.3 & 1115 & 1039.1& -11.61 & 30.48 & -0.14 & 0.22 & 0.09 \\
J182028.37-161030.4 & 31.6 & 1786 & 1771.1 & -11.61 & 31.83 &  0.58 & 0.94 & 0.94 \\
J182041.14-161530.6 & 320.7 & 2015 & 2003.3 & -11.44 & 32.01 & 0.71 & 0.91 & 0.99 \\
J182016.54-161003.0 & 192.1 & 1006 & 1001.8 & -11.54 & 31.90 & 0.99 & 0.01 & 0.99 \\
J054131.61-015232.2 & 272.3 & 1341 & 1337.1 & -11.43 & 30.54 & 0.50 & 0.89 & 0.96 \\
J054148.22-015602.0 & 57.8 & 1636 & 1633.5 & -11.17 & 30.83 & 0.97 & 0.69 & 1.00 \\
J054145.07-015144.6 & 207.9 & 1679 & 1677.7 & -11.24 & 30.75 & 0.84 & 0.94 & 0.99 \\
J064113.15+092611.0 & 528.3 & 1442 & 1418.0 & -11.57 & 31.03  & -0.17 & 0.47 & 0.32 \\
J064058.51+093331.7 & 42.9 & 3089 & 3086.3 & -11.30 & 31.32 & 0.03 & 0.41 & 0.44  \\
J064105.35+093313.3 & 115.4 & 2845 & 2841.9 & -11.39 & 31.24 & -0.02 & 0.43 & 0.42 \\
J053523.47-051850.0 & 277.6 & 1267 & 1262.0 & -11.47 & 30.72  & -0.04 & 0.61 & 0.59 \\
J164055.46-490102.6 & 783.4 & 1574 & 1403.1 & -11.45 & 38.6  & 0.33 & 0.77 & 0.88 \\
J032929.26+311834.6 & 377.2 & 1550 & 1544.2 & -11.08 & 31.59 & 0.43 & 0.83 & 0.93 \\
J015751.17+375257.0 & 421.1 & 2603 & 2583.7 & -11.50 & 30.51 & -0.41 & 0.15 & 0.28 \\
J015751.43+375305.3 & 429.9 & 3971 & 3951.5 & -11.35 & 30.67 & -0.43 & 0.18 & -0.27 \\
J060732.35-061215.5 & 691.8 & 1479 & 1433.2 & -11.49 & 31.15 & 0.26 & 0.81 & 0.89 \\
J060814.11-062558.4 & 394.7 & 2222 & 2209.2 & -11.23 & 31.44 & 0.64 & 0.88 & 0.97 \\
J180438.93-242533.2 & 332.0 & 1227 & 1223.5 & -11.41 & 31.62 & 0.22 & 0.68 & 0.79 \\
J071845.24-245643.8 & 38.4 & 1510 & 1508.2 & -11.59 & 31.55 & 0.04 & 0.51 & 0.54  \\
J180402.88-242140.0 & 252.3 & 1893 & 1882.1 & -11.63 & 31.38 & 0.00 & 0.58 & 0.58 \\
J085932.21-434602.6 & 75.1 & 1675 & 1671.0 & -11.20 & 31.86 & 0.80 & 0.92 & 0.99 \\
J182943.01+010207.1 & 819.2 & 2830 & 2585.7 & -11.29 & 30.36 & 0.25 & 0.76 & 0.85 \\
J034359.71+321403.8 & 516.5 & 2222 & 2211.0 & -11.02 & 30.85 & 0.26 & 0.79 & 0.87 \\
\enddata
\tablenotetext{a}{Units of $log$[ergs cm$^{-2}$ s$^{-1}$]}
\tablenotetext{b}{Units of $log$[$10^{30}$ ergs s$^{-1}$]}
\end{deluxetable}
\clearpage

\begin{landscape}
\begin{deluxetable}{rrrccrrrr}
\tabletypesize{\scriptsize}
\tablecaption{Log of the 17 $\chandra$ observations used.\label{tbl-2}}
\tablewidth{0pt}
\tablehead{
\colhead{Pointing RA} & \colhead{Pointing Dec} &  \colhead{Obs. ID} & \colhead{Target Name} & \colhead{Cluster Age} & \colhead{Cluster Dist.} & \colhead{Expos. Time}  & \colhead{Obs. Start Date} & \colhead{Reference}\\
\colhead{~} & \colhead{~} & \colhead{~} & \colhead{~} &\colhead{[Myr]} & \colhead{[pc]} & \colhead{[ks]} & \colhead{~} & \colhead{~}
}
\startdata
16:27:17.18 & -24:34:39.0 & 635  & Rho Oph. Core      & 0.3      & 130  & 101.87 & Apr 13 2000 & Imanshi \e (2001)\\
16:26:34.20 & -24:23:27.8 & 637  & Rho Oph. A         & 0.3      & 130  & 97.6   & May 15 2000 & Imanshi \e (2001)\\
05:35:19.98 & -05:05:29.8 & 634  & OMC 2/3            & 1        & 450  & 89.17  & Jan 01 2000 & Tsuboi \e (2001)\\
18:20:20.29 & -16:10:44.9 & 6420 & M17 Pointing I     & $<$ 1    & 2100 & 151.36 & Aug 01 2006 & Povich \e (2009)\\
05:41:46.30 & -01:55:28.7 & 1878 & NGC 2024           & 1 $-$ 3  & 380  & 76.43  & Aug 08 2001 & Skinner \e (2003)\\
06:40:58.10 & +09:34:00.4 & 2540 & NGC 2264           & 3        & 760  & 96.97  & Oct 28 2002 & Rebull \e (2006)\\
05:35:15.00 & -05:23:20.0 & 18   & Trapezium Cluster  & 0.8 $-$ 1& 450  & 47.44  & Oct 12 1999 & Garmire \e (2000)\\
16:40:00.10 & -48:51:45.0 & 4503 & RCW 108            & 3        & 1300 & 9.96   & Oct 25 2004 & Wolk \e (2008)\\
03:29:02.00 & +31:20:54.0 & 6436 & NGC 1333           & 1        & 300  & 36.95  & Jul 05 2006 & Winston \e (2010)\\
01:57:39.00 & +37:46:10.0 & 3752 & NGC 752            & 1900     & 400  & 134.08 & Sep 29 2003 & Giardino \e (2008)\\
06:07:49.50 & -06:22:54.7 & 1882 & Monoceros R2       & 6 $-$ 10 & 830  & 98.08  & Dec 12 2000 & Kohno \e (2002)\\
18:04:24.00 & -24:21:20.0 & 977  & NGC 6530           & 2        & 400  & 60.93  & Jun 18 2001 & Damiani \e (2004)\\
07:18:42.80 & -24:57:18.5 & 4469 & NGC 2362           & $~$ 5    & 1500 & 99.14  & Dec 23 2003 & Delgado \e (2006)\\
18:03:45.10 & -24:22:05.0 & 3754 & M8                 & 2        & 1300 & 129.60 & Jul 25 2003 & Arias \e (2006)\\
08:59:27.50 & -43:45:24.0 & 6433 & RCW 36             & 2 $-$ 3  & 700  & 71.30  & Sep 23 2006 & Baba \e (2004)\\
18:29:50.00 & +01:15:30.0 & 4479 & Serpens Cloud Core & $<$ 3    & 260  & 89.58  & Jun 19 2004 & Giardino \e (2007)\\
03:44:30.00 & +32:08:00.0 & 606  & IC 348             & 2        & 320  & 52.96  & Sep 21 2000 & Preibisch \& Zinnecker (2001)\\
\enddata
\end{deluxetable}
\end{landscape}
\clearpage
\setlength{\tabcolsep}{0.068in}
\begin{deluxetable}{clcrcccccccccc}
\tabletypesize{\scriptsize}
\tablecaption{Available infrared photometry for the 30 flare events in the sample.\label{tbl-3}}
\tablewidth{0pt}
\tablehead{
\colhead{$\chandra$ Obs ID} & \colhead{ANCHORS ID} & \colhead{YSO class} &
\colhead{$A_{K_s}$} & \colhead{$J$} & \colhead{$H$} & \colhead{$K_s$} &
\colhead{[3.6 $\micron$]} & \colhead{[4.5 $\micron$]} &
\colhead{[5.8 $\micron$]} & \colhead{[8.0 $\micron$]} & \colhead{[24 $\micron$]}\\
\colhead{~}& \colhead{[CXOANC ...]} & \colhead{~}
}
\startdata
635 & J162727.0-244049 & 1 & 4.80 & \nodata & 13.52  & 9.74  & 6.71 1 & 5.48 & 4.57 & 3.80 & \nodata \\
635 & J162622.5-242251 & 2 & 0.28 & 16.45 & 15.09 & 13.94 & 12.39 & 11.55 & 10.86 & 9.82 & 5.78 \\
635 & J162715.5-242639 & 2 & 3.60 & 17.42 & 13.46 & 10.79  & 8.21  & 7.43  & 6.80  & 6.24  & 3.49 \\
637 & J162624.1-242447 & 2 & 1.69 & 11.12 & 8.72 & 7.32  & 5.63  & 4.89  & 4.41 & 3.61  & 1.23  \\
634 & J053456.8-051133 & 2 & 0.00 & 10.90 & 10.35 & 10.15 & 8.92 & 8.61 & 8.19 & 7.46 & \nodata \\
634 & J053525.7-050949 & 3 & 0.00 & 11.07 & 10.43 & 10.11 & 9.69 & 9.53 & 9.30 & 8.35 & \nodata \\
634 & J053629.6-045359 & \nodata & \nodata & 12.66 & 11.97 & 11.74 & \nodata & \nodata & \nodata & \nodata & \nodata \\
6420 & J182028.4-161030	& \nodata & \nodata & 13.77 & 12.41 & 11.98 & \nodata & \nodata & \nodata & \nodata & \nodata \\
6420 & J182041.1-161530	& 2 & \nodata & 15.48 & 13.52 & 12.61 & \nodata & \nodata & \nodata & \nodata & \nodata \\
6420 & J182016.5-161003	& 2 & \nodata & 16.11 & 15.04 & 13.74 & \nodata & \nodata & \nodata & \nodata & \nodata \\
1878 & J054131.6-015232	& 2 & 0.53 & 14.35 & 13.14 & 12.53 & 11.68 & 11.31 & 10.37 & 9.11 & \nodata \\
1878 & J054148.2-015602	& 3 & 4.39 & \nodata & 13.82 & 11.20 & 9.51 & 9.11 & 7.93 & \nodata & \nodata \\
1878 & J054145.1-015144	& 3 & 2.11 & 14.23 & 11.61 & 10.29 & 9.48 & 9.24 & 9.14 & 9.31 & \nodata \\
2540 & J064113.2+092610	& 2 & \nodata & 11.61 & 10.81 & 10.34 & \nodata & \nodata & \nodata & \nodata & \nodata \\
2540 & J064058.5+093331	& \nodata & \nodata & 10.22 & 10.23 & 10.17 & \nodata & \nodata & \nodata & \nodata & \nodata \\
2540 & J064105.4+093313	& 2 & \nodata & 12.48 & 11.85 & 11.64 & \nodata & \nodata & \nodata & \nodata & \nodata \\
18 & J053523.5-051849 & 3 & 0.26 & 11.99 & 11.12 & 10.79 & 10.61 & 10.26 & 8.98 & \nodata & \nodata \\
4503 & J164055.5-490102	& 3 & 0.67 & 13.11 & 11.88 & 11.42 & 11.21 & 11.09 & 10.90 & 11.03 & \nodata \\
6436 & J032929.2+311834	& 2/3\tablenotemark{a} & 0.65 & 12.58 & 11.40 & 10.98 & 10.67 & 10.65 & 10.52 & 10.05 & 4.71 \\
3752 & J015751.2+375257	& 3 & \nodata & 12.01 & 11.38 & 11.18 & \nodata & \nodata & \nodata & \nodata & \nodata \\
3752 & J015751.4+375305	& 3 & \nodata & 10.11 & 9.70 & 9.57 & \nodata & \nodata & \nodata & \nodata & \nodata \\
1882 & J060732.4-061215	& \nodata & \nodata & 14.42 & 13.39 & 13.04 & \nodata & \nodata & \nodata & \nodata & \nodata \\
1882 & J060814.1-062558	& 2 & 0.43 & 12.51 & 11.10 & 10.02 & 8.45 & 7.96 & 7.32 & 6.70 & 3.11 \\
977 &  J180438.9-242533 & 3 & \nodata & 14.26 & 13.45 & 13.12 & \nodata & \nodata & \nodata & \nodata & \nodata \\
4469 & J071845.2-245643 & 3 & \nodata & 15.06 & 14.47 & 14.24 & \nodata & \nodata & \nodata & \nodata & \nodata \\
3754 & J180402.9-242139 & 3 & \nodata & 11.98 & 11.27 & 11.06 & \nodata & \nodata & \nodata & \nodata & \nodata \\
6433 & J085932.2-434602 & 2 & \nodata & 14.42 & 12.61 & 11.85 & \nodata & \nodata & \nodata & \nodata & \nodata \\
4479 & J182943.0+010207 & 3 & 0.54 & 12.86 & 11.74 & 11.35 & 11.03 & 10.98 & 10.95 & 10.89 & \nodata \\
606 & J034359.7+321403 & 1 & 0.12 & 12.31 & 11.40 & 11.06 & 13.77 & 12.99 & 12.20 & 11.19 & \nodata \\
\enddata
\tablenotetext{a}{This source was identified as having a transition disk via the presence of excess 24 $\micron$ emission}
\tablecomments{IRAC photometry is all class A with average error of 0.01 magnitudes.  Errors on the 2MASS photometry ranges from 0.03-0.1 magnitudes}
\end{deluxetable}
\clearpage
\begin{deluxetable}{crccccccccc}
\setlength{\tabcolsep}{0.068in}
\tabletypesize{\scriptsize}
\tablecaption{Calculated Flare Data \label{tbl-4}}
\tablewidth{0pt}
\tablehead{
\colhead{Obs ID} & \colhead{ANCHORS ID} & \colhead{$\tau_{LC}$} & \colhead{Max $T_{OBS}$} &
\colhead{$T_{PK}$} & \colhead{$\zeta$} & \colhead{$\Delta~\zeta$} &
\colhead{Loop Length} & \colhead{$n_e$} & \colhead{B} & \colhead{Tot Flare Energy}\\
\colhead{~} & \colhead{[CXOANC ...]} & \colhead{[$ks$]} & \colhead{[MK]} &
\colhead{[MK]} & \colhead{~} & \colhead{~} & \colhead{[$10^{10}$ cm]} & \colhead{[$10^{10}~cm^{-3}$]} 
& \colhead{[$G$]} & \colhead{log[ergs]} 
}
\startdata
\multicolumn{11}{l}{Class I YSOs} \\
635 & J162727.0-244049 & 15.0 & 65 & 160 & 1.64 & 1.07 & 26.4 (26 $-$ 131) & 3.7 & 45 & 34.96 \\
606 & J034359.7+321403 & 15.8 & 38 & 84 & 0.53 & 0.28 & 9.0 ($<$ 15) & 18.8 & 74 & 35.25 \\
\multicolumn{11}{l}{Class II YSOs} \\
635 & J162622.5-242251 & 6.5 & 38 & 85 & 0.60 & 0.34 & 4.4 ($<$ 6.7) & 38.1 & 106 & 34.56 \\
635 & J162715.5-242639 & 10.5 & 72 & 182 & 1.79 & 0.65 & 19.7 (17 $-$ 20) & 8.4 & 73 & 35.56 \\
637 & J162624.1-242447 & 12.9 & 85 & 224 & 3.73 & 5.06 & 26.9 (\nodata) & 2.3 & 42 & 34.99 \\
634 & J053456.8-051133 & 11.5 & 23 & 46 & 0.58 & 0.37 & 5.4 ($<$ 8.7) & 2.9 & 68 & 35.00 \\
2540 & J064113.2+092610 & 9.5 & 50 & 118 & 2.01 & 3.10 & 14.4 (\nodata) & 4.3 & 42 & 35.68 \\
6420 & J182041.1-161530 & 12.0 & 126 & 358 & 0.37 & 0.23 & 4.5 ($<$ 17) & 234.0 & 539 & 37.01 \\
6420 & J182016.5-161003 & 20.7 & 91 & 243 & 1.39 & 0.73 & 43.6 (27 $-$ 45) & 8.2 & 83 & 37.04 \\
1878 & J054131.6-015232 & 10.8 & 50 & 118 & 1.51 & 1.56 & 16.3 (\nodata) & 6.9 & 53 & 35.37 \\
2540 & J064105.4+093313 & 23.5 & 36 & 78 & 0.64 & 0.21 & 16.6 (7.6 $-$ 21) & 5.1 & 37 & 36.13 \\
1882 & J060814.1-062558 & 3.5 & 61 & 149 & 0.78 & 0.45 & 4.2 (0.13 $-$ 5.5) & 16.5 & 293 & 35.85 \\
6433 & J085932.2-434602 & 9.0 & 123 & 346 & 2.36 & 1.43 & 23.3 (18 $-$ 24) & 20.8 & 158 & 36.66 \\
COUP 1246 & \nodata & 22.5 & 126 & 357 & 1.026 & 0.21 & 50.0 (43 $-$ 55) & 2.0 & 50 & \nodata \\
\multicolumn{11}{l}{Class III YSOs} \\
1878 & J054148.2-015602 & 8.0 & 93 & 249 & 0.87 & 0.49 & 13.3 (3 $-$ 17) & 14.0 & 110 & 35.59 \\
1878 & J054145.1-015144 & 19.0 & 40 & 91 & 0.58 & 0.43 & 12.8 ( $<$ 21) & 12.0 & 62 & 35.59 \\
4503 & J164055.5-490102 & 6.4 & 39 & 88 & 0.41 & 0.45 & 1.9 ($<$ 6.2) & 563.0 & 415 & 36.53 \\
18 & J053523.5-051849 & 2.5 & 32 & 68 & 0.4 & 0.12 & 0.6 ( $<$ 1.2) & 990.0 & 485 & 35.47 \\
3752 & J015751.2+375257 & 7.5 & 27 & 57 & 0.49 & 0.213 & 2.9 ($<$ 5) & 106.0 & 145 & 35.53 \\
3752 & J015751.4+375305 & 7.3 & 30 & 63 & 0.62 & 0.30 & 4.4 (0.07 $-$ 6) & 35.4 & 88 & 35.27 \\
634 & J053525.7-050949 & 5.0 & 21 & 41 & 3.89 & 5.35 & 10.2 (\nodata) & 18.4 & 51 & 34.97 \\
4469 & J071845.2-245643 & 8.8 & 46 & 107 & 0.44 & 0.20 & 3.8 ($<$ 7) & 153.0 & 237 & 36.54 \\
977 & J180438.9-242533 & 4.8 & 59 & 143 & 1.11 & 0.88 & 7.0 ($<$ 8) & 129.0 & 253 & 36.33 \\
6436 & J032929.2+311834 & 1.6 & 79 & 205 & 1.32 & 0.63 & 3.0 (2 - 3.2) & 133.0 & 307 & 35.32 \\
3754 & J180402.9-242139 & 5.0 & 36 & 80 & 0.63 & 0.29 & 3.5 (0.53 $-$ 5) & 242.0 & 259 & 36.02 \\
4479 & J182943.0+010207 & 3.5 & 58 & 140 & 1.92 & 1.35 & 5.8 (3 $-$ 6) & 44.6 & 147 & 35.19 \\
\multicolumn{11}{l}{Unknown Class} \\
6420 & J182028.4-161030 & 7.5 & 82 & 214 & 0.72 & 0.71 & 10.0 ($<$ 15) & 38.4 & 169 & 36.50 \\
2540 & J064058.5+093331 & 20.5 & 93 & 248 & 1.41 & 0.40 & 43.9 (37 $-$ 45)  & 1.5 & 36 & 36.33 \\
1882 & J060732.4-061215 & 17.0 & 36 & 80 & 0.52 & 0.24 & 9.1 ($<$ 15) & 32.6 & 95 & 36.07 \\
634 & J053629.6-045359 & 22.5 & 36 & 78 & 0.84 & 0.36 & 20.5 (10 $-$ 25) & 3.1 & 29 & 35.16 \\
\enddata
\tablecomments{The flare loop lengths are presented with their uncertainty ranges in parentheses}
\end{deluxetable}


\end{document}